\documentclass[lettersize,journal]{IEEEtran}
\usepackage{amsmath,amsfonts}
\usepackage{algorithmic}
\usepackage{algorithm}
\usepackage{array}
\usepackage[caption=false,font=normalsize,labelfont=sf,textfont=sf]{subfig}
\usepackage{textcomp}
\usepackage{stfloats}
\usepackage{url}
\usepackage{verbatim}
\usepackage{graphicx}
\usepackage{cite}
\usepackage{float}
\usepackage{cleveref}
\usepackage{multirow}
\crefname{figure}{Fig.}{Figs.}

\usepackage{booktabs}
\usepackage{latexsym}
\usepackage{amsfonts}
\usepackage{amsmath}
\usepackage{xcolor}
\usepackage{colortbl}
\usepackage{epsfig}
\usepackage{xspace}
\usepackage{graphicx}
\usepackage{paralist}
\usepackage{enumerate}
\usepackage[color,matrix,arrow,all]{xy}
\usepackage{comment}
\usepackage{booktabs}
\usepackage{balance}
\usepackage{stmaryrd}
\usepackage{pifont}
\usepackage{hhline}
\usepackage{listings}
\usepackage{array}
\usepackage{float}
\usepackage[flushleft]{threeparttable}

\usepackage{mathrsfs}
\usepackage{makecell}
\usepackage{xparse}
\usepackage{wrapfig}

\usepackage{epsfig}
\usepackage{multirow}
\usepackage{url}

\usepackage{multirow}
\usepackage{graphicx}

\usepackage{tcolorbox}

\usepackage{listings}
\usepackage{framed}
\usepackage{xcolor}
\usepackage{color}
\usepackage{algorithm}
\usepackage{algorithmic}
\usepackage[algo2e]{algorithm2e} 
\usepackage[export]{adjustbox}% http://ctan.org/pkg/adjustbox

\usepackage{tikz}

\newcommand{\eat}[1]{}

\colorlet{shadecolor}{gray!20}

\definecolor{shadecolor}{RGB}{220,220,220}

\definecolor{inputcolor}{RGB}{255,139,35}
\definecolor{outputcolor}{RGB}{120,212,252}
\definecolor{embedcolor}{RGB}{254,127,156}
\definecolor{maskcolor}{RGB}{122,128,255}
\definecolor{ecolor}{RGB}{58,149,54}

\definecolor{highcolor}{RGB}{255,153,153}
\definecolor{midcolor}{RGB}{255,204,204}
\definecolor{lowcolor}{RGB}{204,229,255}

\usetikzlibrary{shapes,snakes}
\usetikzlibrary{calc}

\definecolor{green}{RGB}{0,128,0}

\definecolor{yellow}{RGB}{255,200,18}

\newcommand{\at}[1]{\protect\ensuremath{\mathsf{#1}}}

\newcommand{\stab}{\vspace{1.2ex}\noindent}

\newtheorem{example}{Example}

\newcommand{\etal}{{\emph{et al}.\thinspace}}

\newcommand{\bi}{\begin{itemize}}
\newcommand{\ei}{\end{itemize}}

\newcommand{\be}{\begin{enumerate}}
\newcommand{\ee}{\end{enumerate}}
\newcommand{\beqn}{\begin{eqnarray*}}
\newcommand{\eeqn}{\end{eqnarray*}}

\newcommand{\stitle}[1]{\stab\noindent{\bf #1}}
\newcommand{\etitle}[1]{\vspace{1mm}\noindent{\underline{\em #1}}}

\newcommand{\ie}{\textit{i.e.,}\xspace}
\newcommand{\eg}{\textit{e.g.,}\xspace}

\newcommand{\sys}{{\at {ChartEditor}}\xspace}

\makeatletter
    \newcommand\figcaption{\def\@captype{figure}\caption}
    \newcommand\tabcaption{\def\@captype{table}\caption}
\makeatother

\tikzstyle{mybox} = [draw=black, fill=black!5, thick,
   rectangle, rounded corners, inner sep=0pt, inner ysep=6pt]
\tikzstyle{fancytitle} =[fill=black, text=white]

\NewDocumentCommand{\nan}{ mO{} }{\textcolor{blue}{\textsuperscript{\textit{Nan}}\textsf{\textbf{\small[#1]}}}}

\NewDocumentCommand{\yuyu}{ mO{} }{\textcolor{green}{\textsuperscript{\textit{Yuyu}}\textsf{\textbf{\small[#1]}}}}

\NewDocumentCommand{\vica}{ mO{} }{\textcolor{blue}{\textsuperscript{\textit{Weikai}}\textsf{\textbf{\small[#1]}}}}

\begin{document}

\title{ChartEditor: A Human-AI Paired Tool for Authoring Pictorial Charts}

\author{
    Siyu Yan,
    Tiancheng Liu,    
    Weikai Yang$^*$,
    Nan Tang,
    Yuyu Luo$^*$
    \thanks{
    S. Yan, T. Liu, W. Yang, N. Tang, and Y. Luo tare with the Hong Kong University of Science and Technology (Guangzhou). 
    E-mail: \{syan195, tcliu767\}@connect.hkust-gz.edu.cn.
    \{weikaiyang, nantang, yuyuluo\}@hkust-gz.edu.cn.
    Weikai Yang and Yuyu Luo are the corresponding authors.\\  }
    % <-this % stops a space
}

\markboth{Journal of \LaTeX\ Class Files,~Vol.~14, No.~8, August~2021}%
{Shell \MakeLowercase{\textit{et al.}}: A Sample Article Using IEEEtran.cls for IEEE Journals}

\maketitle

\begin{abstract}
Pictorial charts are favored for their memorability and visual appeal, offering a more engaging alternative to basic charts.
However, their creation can be complex and time-consuming due to the lack of native support in popular visualization tools like Tableau.
While AI-generated content (AIGC) tools have lowered the barrier to creating pictorial charts, they often lack precise design control. 
To address this issue, we introduce \sys, a human-AI paired tool that transforms basic charts into pictorial versions based on user intent.
\sys decomposes chart images into visual components and organizes them within a hierarchical tree. %  that enables fine-grained design control
Based on this tree, users can express their intent in natural language, which is then translated into modifications to the hierarchy.
In addition, users can directly interact with and modify specific chart components via an intuitive interface to achieve fine-grained design control.
% In addition, an interactive interface allows users to manipulate specific chart components directly.
A user study demonstrates the effectiveness and usability of \sys in simplifying the creation of pictorial charts.
\end{abstract}


%!TEX root = ../main.tex
\section{Introduction}
\begin{figure*}[t]
    \centering
    \vspace{-1em}
\includegraphics[width=\textwidth]{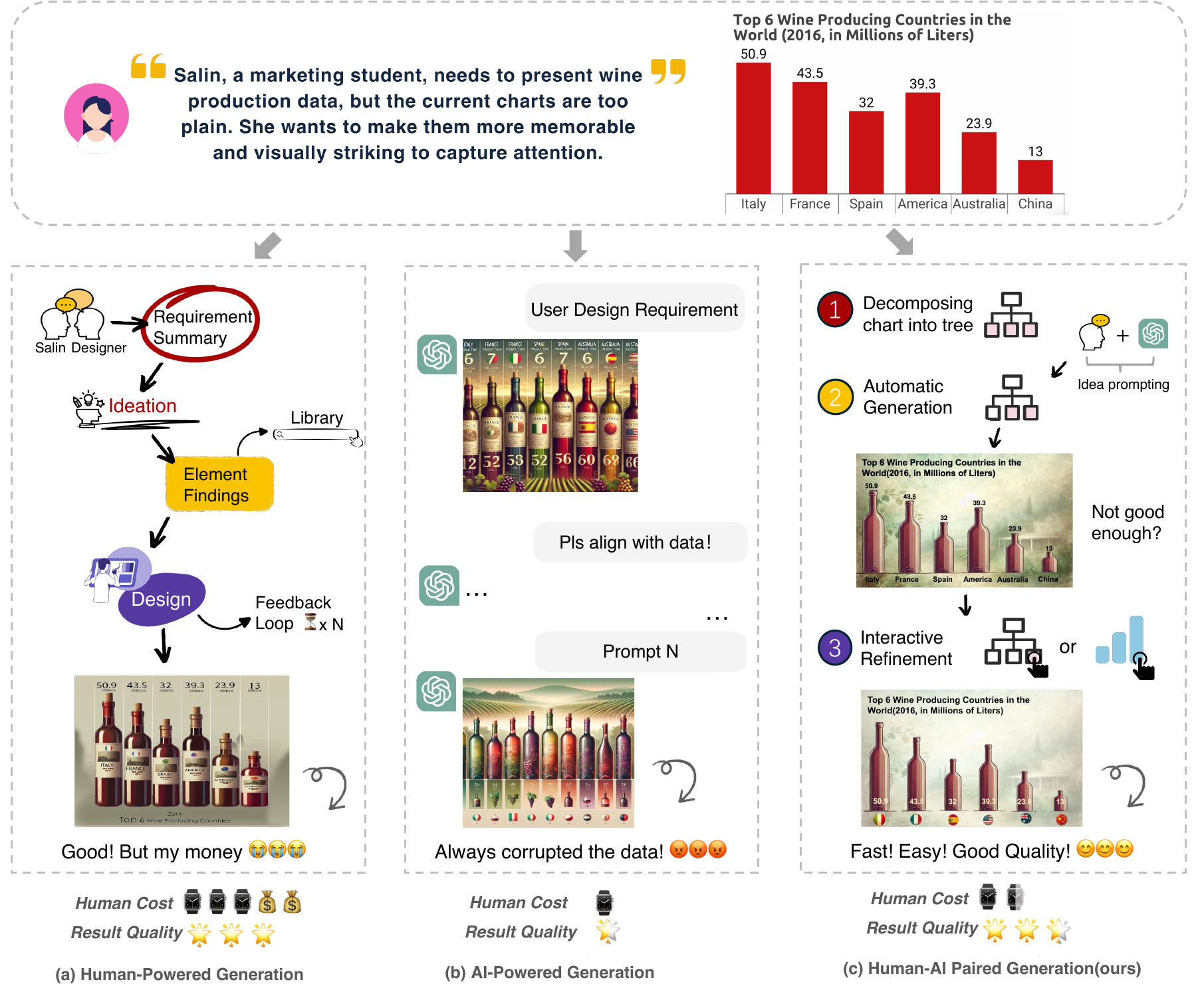}
    \caption{Comparison of Pictorial Chart Generation Methods. (a) Manual creation with Adobe Illustrator requires extensive collaboration and iterative revisions to produce high-quality charts. (b) AI-generated methods offer quick and visually appealing charts but distort the original data. (c) \sys balances automation and user control, enabling efficient, accurate, and customizable chart generation.}
    \label{fig:teaser}
\end{figure*}

Basic charts, such as bar charts and pie charts, rely on simple geometric shapes to effectively convey data trends and comparisons~\cite{4376133,DBLP:journals/vldb/QinLTL20,DBLP:conf/icde/LuoQ0018}. While functional and widely used, these charts often fall short in engaging audiences or making the information visually memorable and contextually meaningful~\cite{rolfes2021interpretation}.
Pictorial charts overcome this limitation by incorporating contextually relevant images or icons.
For example, as shown at the top of \cref{fig:teaser}, Salin, a marketing student, needed to present wine production data in a more engaging and visually striking way. By replacing standard chart elements with meaningful icons, such as wine bottles, she was able to create a chart that immediately captured attention and conveyed the information more memorably.
These visual elements provide immediate, intuitive cues that enhance both the aesthetic appeal and the retention of information. As a result, pictorial charts are particularly popular in scenarios requiring quick comprehension, such as media, education, and presentations targeting broad and diverse audiences~\cite{zhao2023stories}.

However, creating pictorial charts is significantly more complex than generating basic charts due to the lack of native support in popular visualization tools like Tableau and PowerBI.
Traditionally, users employ \textbf{human-powered tools} such as Adobe Illustrator to craft pictorial charts manually.
As shown in \cref{fig:teaser}(a), users start by quickly sketching a rough design based on their experience.
Next, they search for appropriate materials and manually adjust them to bring the design to life.
Depending on the visual results, users may need to revisit the design plan or replace materials.
While these tools offer flexibility and control over the design, they require extensive manual adjustments, demanding a lot of time and effort to achieve visual appeal and accurate data representation.
This makes the creation of pictorial charts particularly challenging and inaccessible for users without extensive design expertise.

% As shown in \cref{fig:reference-label}, users often attempt to create pictorial charts by {\em transforming} basic charts into pictorial versions, replacing standard elements with contextually relevant icons or images.

% \begin{figure}[H]
%     \centering
%     \includegraphics[width=0.7\linewidth]{figure/1-reference chart.pdf}
%     \caption{Transforming a basic bar chart into a pictorial chart by replacing bars with rocket icons.}
%     \label{fig:reference-label}
    
% \end{figure}

With the advancements in generative AI, users can now utilize AIGC tools like DALL·E 3~\cite{openai2024dalle3} or MidJourney~\cite{midjourney_2023} to generate pictorial charts. Although these \textbf{AI-powered tools} greatly reduce the manual effort, they often lack the precision and control necessary for accurate data representation and may require further refinement by users~\cite{DBLP:journals/vi/YeHHWXLZ24,DBLP:journals/pvldb/XieLLT24,DBLP:journals/tvcg/LiLFZLL24,DBLP:journals/corr/abs-2406-03753}.
As shown in \cref{fig:teaser}(b), the AI-generated pictorial chart fails to align visual elements with the original data, leading to user frustration due to insufficient support for manual adjustments.

\IEEEpubid{0000--0000/00\$00.00~\copyright~2021 IEEE}

To address these limitations and balance the strengths of human-powered and AI-powered tools, we present \sys, a \textbf{human-AI paired tool} that combines the best of both worlds.
As depicted in \cref{fig:teaser}(c), \sys begins by automatically generating a pictorial chart based on user intent expressed in natural language. The system then supports iterative refinement for fine-grained customization. Specifically, \sys decomposes a basic chart into its visual elements (e.g., marks, axes) and organizes them into a chart tree that reflects the hierarchical structure of the chart (\cref{fig:teaser}(c)-\ding{172}). Each node in the tree represents a specific visual component, maintaining the relationships between these elements.
Second, users can then express their design intent in natural language, which is translated into operations to the chart tree (\cref{fig:teaser}(c)-\ding{173}).
In addition, users can directly interact with and modify specific chart components via an intuitive interface to achieve fine-grained design control ,\eg replacing the bars using bottles and adding a background image (\cref{fig:teaser}(c)-\ding{174}).
By combining automation with interactive refinement, \sys empowers users to create accurate and visually appealing pictorial charts with far less time and effort compared to traditional methods.

% Specifically, all modifications made to nodes or paths within the hierarchical tree are directly synchronized with their corresponding elements in the chart.
% This linkage ensures that any changes to the tree—whether adding, removing, or altering nodes and paths—are automatically reflected in the visual representation of the pictorial chart.
% Therefore, users can perform multi-granularity edits with high precision, adjusting both broad structural elements and fine details seamlessly.
% As shown in Figure~\ref{fig:teaser}(c)-\ding{174}, if the initial mapping between the chart tree and the chart components is inaccurate, users can select specific nodes or paths to make precise adjustments by refining the tree structure or directly manipulating chart elements.
% Thus, \sys empowers users to create accurate and visually appealing pictorial charts without the extensive time and effort typically required by traditional methods.

% We conducted a quantitative experiment to and a controlled user study to evaluate the effectinvess of \sys.

\IEEEpubidadjcol % 双列排版显示正确脚标

In summary, our contributions include:
\bi 
\item We develop \sys, that leverages both human input and AI capabilities, allowing users to generate and refine pictorial charts through natural language and interactive manipulation. (Sections~\ref{system} and~\ref{sec:usage})

\item We propose the Chart Tree framework for precise and consistent modification of pictorial charts. To support this framework, we curate the ChartSS dataset with 59,693 annotated charts and train a chart segmentation model to segment and organize visual elements of basic charts for integration into the Chart Tree. (Sections~\ref{sec:tree} and~\ref{sub:decompose})

\item We conduct a quantitative evaluation to demonstrate the effectiveness of our curated dataset and the chart decomposition method and a user study to evaluate the usability of \sys. (Sections~\ref{sec:quan} and~\ref{sec:userstudy})

\ei

\section{Related work}
\subsection{Pictorial Chart}

Compared to basic charts, pictorial charts utilize pictorial objects, such as realistic photographs and abstract pictograms, to enhance memorability and user engagement~\cite{bateman2010useful,alebri2024visualisations}.
Borkin~\etal\cite{borkin2013makes} studied the memorability of visualizations and confirmed that the inclusion of pictorial objects would enhance memorability.
Moving beyond memorability, Borkin~\etal\cite{borkin2015beyond} found that appropriate use of pictograms will not hinder understanding but rather enhance recognition.
Similarly, Alebri~\etal\cite{alebri2024visualisations} verified that adding semantically related icons, such as flags next to country names, can enhance perceived engagement.
However, some researchers also pointed out that the introduction of irrelevant pictorial objects can be distracting and confusing~\cite{haroz2015isotype,borgo2012empirical}.
For example, Haroz~\etal\cite{haroz2015isotype} observed that superfluous pictographs and label images can confuse and distract readers.
Borgo~\etal\cite{borgo2012empirical} also noted that pictorial charts could negatively impact visual search tasks, especially when the readers are not familiar with the pictograms used.
Therefore, it is crucial to maintain semantic relevance between the pictorial objects and the chart's underlying narrative, which will better engage readers.

\subsection{Chart Deconstruction}

Chart deconstruction aims to decompose charts and extract the underlying data from them.
Existing methods can be classified into two categories based on the chart format they process: vector graphics charts and rasterized charts~\cite{DBLP:conf/icde/ChaiLFL20,DBLP:journals/tkde/ChaiLFL21,DBLP:journals/pacmmod/LuoZ00CS23}.

Deconstruction methods for vector graphics charts usually leverage the inherent benefits of the format, such as high resolution, clear structure, and precise element positioning and sizing, thus enhancing the quality of deconstruction.
For example, Harper and Agrawala~\cite{harper2014deconstructing} parsing the SVG tree from a D3-generated chart to extract the underlying data, the visual marks, and the mappings between the data and the mark attributes.
This methodology allows users to restyle D3 visualizations without manually revising JavaScript code.
Later, they extend it to extract additional structure, such as axis orientation and mark groups, which achieves better restyling~\cite{harper2017converting} and facilitates visualization search~\cite{hoque2019searching}.
However, these methods are limited to charts generated with D3.
To encompass a broader range of vector graphics charts, ChartDetective~\cite{masson2023chartdetective} allows users to interactively select marks and axes for better chart deconstruction.
Mystique~\cite{chen2023mystique} parses the SVG tree to identify reusable layout components for further reuse.
However, in many application scenarios, only rasterized charts are available, which limits the applicability of these vector graphics-based methods.

To tackle this issue, substantial efforts have also been made to reconstruct rasterized charts.
For example, Revision~\cite{savva2011revision} first classifies chart type using a support vector machine and subsequently extracts marks and data from an input chart image.
ChartOCR~\cite{luo2021chartocr} supports extracting data from different chart types, including bar charts, pie charts, and line charts.
This is achieved by detecting key points of the visual marks, identifying the chart type, and then translating these key points into numerical values.
ChartDETR~\cite{xue2023chartdetr} and ChartReader~\cite{cheng2023chartreader} utilize transformer-based models to detect the key points of chart components for component detection and achieve better results.
In addition to these fully automatic methods, some efforts incorporate human feedback in the deconstruction process to achieve better results when automatic approaches fall short.
For example, Poco~\etal\cite{poco2017extracting} allowed users to specify legend regions to enhance the accuracy in recovering color mappings.
ChartSense~\cite{jung2017chartsense} employs a convolutional neural network to classify chart types and provides a user interface to interactively extract marks and data.
Compared to these methods, we construct a dataset designed for chart semantic segmentation and use it to fine-tune a Mask2Former model, which produces higher-quality segmentation masks.
In addition, these visual elements are organized into a hierarchal chart tree to support the adjustment at different levels of granularity.
% Our approach aligns with these practices: we support rasterized charts to broaden the applicability of our method and offer users the flexibility to modify automatic deconstruction results.

% rule-based detection:\cite{balaji1812chart},\cite{gao2012view},\cite{poco2017reverse}, have traditionally been the mainstream approach for addressing the problem of chart element extraction. These methods use color continuity search and edge detection to identify the original components. Predefined rules are then applied to eliminate incorrect candidates. However, these methods heavily rely on manually crafted rules and predefined features, making them effective only for specific types of charts and resulting in poor generalizability.

% \subsection{Styling transfer }
\subsection{Pictorial Chart Authoring}
% \ysy{Related Work: reviewer 2 suggest adding a more in-depth comparison with existing research. For instance, in section 2.2, since there are numerous works on reconstructing rasterized charts, it would be helpful to clarify how your approach differs from or improves upon them.}
Recognizing the benefits of pictorial charts, many researchers have been exploring how to efficiently generate high-quality pictorial charts.
Since pictorial objects are the most important components in pictorial charts, some efforts are devoted to facilitating the design process of pictorial objects.
For example, DataQuilt\cite{zhang2020dataquilt} leverages computer vision techniques to extract and convert real image content into pictorial objects.
MetaGlyph~\etal\cite{ying2022metaglyph} allows users to design metaphoric glyphs based on semantic inputs.
In addition to designing pictorial objects, some tools facilitate the generation of pictorial charts by transferring styles from existing examples.
For example, Retrieve-Then-Adapt~\cite{qian2020retrieve} supports generating proportional-related pictorial charts by first retrieving similar examples from their library and then imitating them.
Chen~\etal\cite{chen2019towards} extracted extensible timeline templates from examples to generate new timeline infographics.
While these methods yield promising results, they are limited to a few visualization types and rely heavily on the quality of examples used.
To address these limitations, Vistylist~\cite{shi2022supporting} automatically extracts visual styles from the source visualizations and allows users to interactively apply them to target data.
This enables a more expressive and faithful representation.
Recently, diffusion models have been adopted to generate pictorial charts based on user intent.
For example, viz2viz~\cite{wu2023viz2viz} first applies mark-level transformations to convert marks into pictorial objects and then applies a chart-level transformation to synthesize a cohesive chart.
However, it does not provide a user-friendly GUI to allow users
to examine intermediate results and directly manipulate them.
ChartSpark~\cite{xiao2023spark} generates foreground and/or background based on the input chart and text prompt, which streamlines the creation of pictorial charts.
A GUI is also provided to assist users in refining the generated charts.
However, these methods are only applicable to charts in vector graphics format to accurately replace visual marks with the generated pictorial objects.
In contrast, \sys is designed to help users without professional design skills to efficiently create pictorial charts based on the rasterized version.

\begin{figure*}[t!]
    \centering
    \begin{minipage}{\textwidth}
        \centering
       \includegraphics[width=0.9\textwidth]{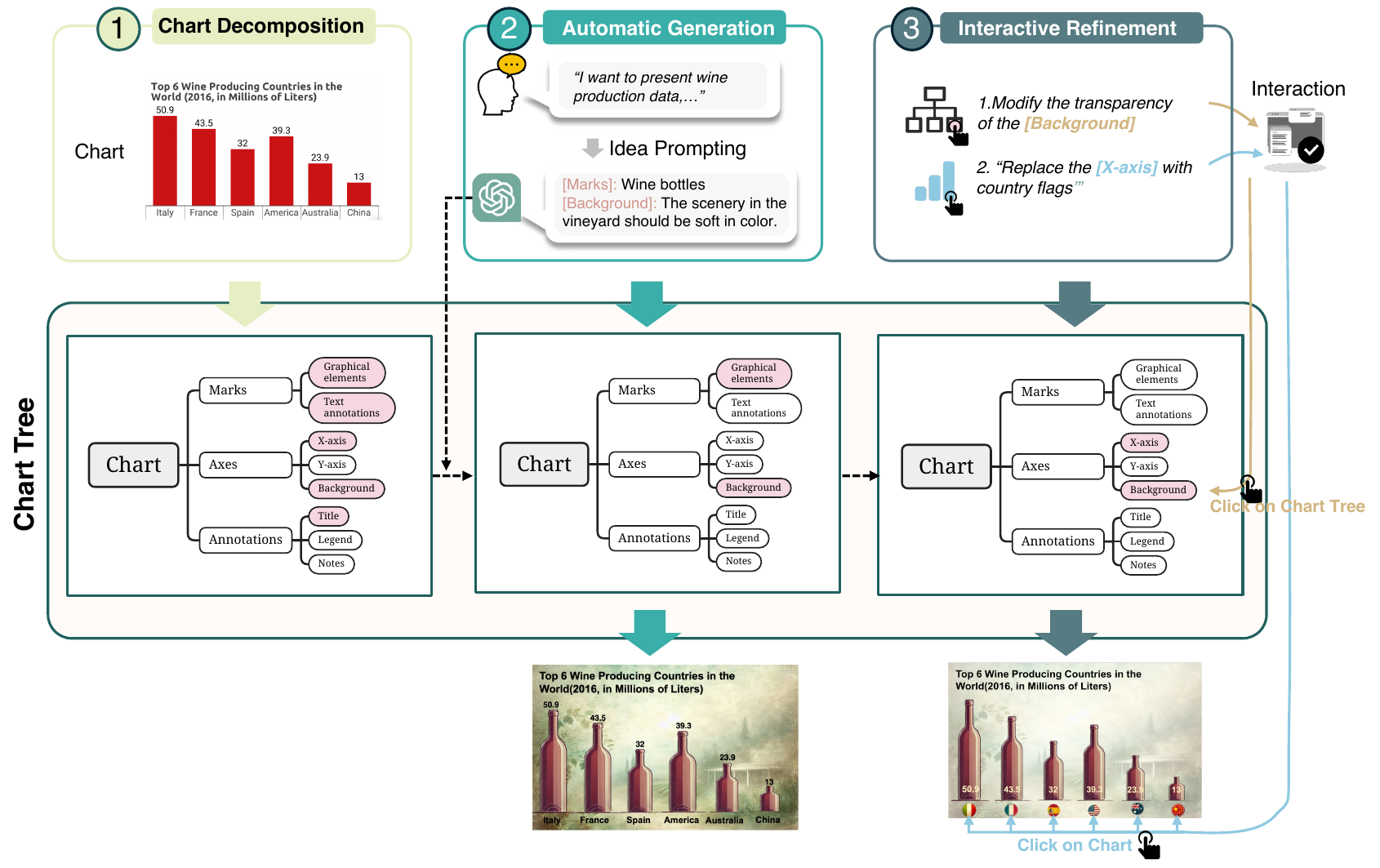}
        \caption{\sys transforms a basic chart into its pictorial version in three steps. 
        {\em (1) Chart Decomposition:} The input chart is broken down into fundamental visual elements, organized within a hierarchical ``Chart Tree'' to enable structured editing. 
        {\em (2) Automatic Generation:} Users provide prompts, and AI generates contextually relevant pictorial elements, such as icons and background elements, that integrate with the basic chart structure. 
        {\em (3) Interactive Refinement:} Users refine the chart by directly modifying components within the Chart Tree or the chart image itself, ensuring precise adjustments and maintaining data integrity.
        The highlighted parts in the Chart Tree indicate components that have been identified or modified during the respective step.
        }
        \label{fig:framework}

    \end{minipage}
\end{figure*}

\section{ChartEditor}
\label{system}

In this section, we first present an overview of \sys (Section~\ref{sub:overview}). 
% Next, we detail the design of the chart tree, which organizes chart elements to enable precise and consistent editing (Section~\ref{sec:tree}). 
Next, we detail the core component of \sys, the chart tree, which organizes visual elements to enable precise and consistent modifications (Section~\ref{sec:tree}).
We then describe the process of constructing a chart tree from a given chart image (Section~\ref{sub:decompose}), automatically modifying the chart tree to generate an initial pictorial chart based on high-level user intent (Section~\ref{sub:autosys}), and interactively refining the charts (Section~\ref{sub:intersys}) through low-level adjustments.
% We then describe the process of decomposing a chart image into fundamental elements that are integrated into the chart tree structure (Section~\ref{sub:decompose}).
% Finally, we explain the automatic generation (Section~\ref{sub:autosys}) and interactive refinement (Section~\ref{sub:intersys}) processes.

\subsection{System Overview}
\label{sub:overview}

\sys is a human-AI paired tool for transforming basic charts into pictorial charts.
The workflow of \sys framework is illustrated in \cref{fig:framework}, which consists of three modules: {\em chart decomposition}, {\em automatic generation}, and {\em interactive refinement}.

\stitle{Chart Decomposition.} 
To enable precise and structured editing, \sys begins by decomposing the input chart image into its visual components, such as marks, axes, and annotations. These elements are organized into a hierarchical \textbf{chart tree}.
This decomposition is essential for enabling targeted, fine-grained edits on the visual elements of interest.
% ensuring that users have a clear and manageable framework for customizing specific chart elements.
% The chart tree offers two key benefits:
% (1) it provides users with a clear visualization of the relationships between different chart components, and
% (2) it simplifies the process of isolating and modifying individual elements within the chart. 
% This decomposition step is essential for enabling targeted, fine-grained edits, ensuring that users have a clear and manageable framework for customizing specific chart elements.

\stitle{Automatic Generation.}
In this module, \sys uses idea prompting to interpret high-level user intent, which may be vague or conceptual, and translate them into predefined modifications on the chart tree.
% For example, users can prompt the system to replace standard chart elements (like bars) with AI-generated pictograms or to customize backgrounds.
For example, users can prompt the system ``\textit{I want to present wine
production data}.'' Then, the system will replace the bars in the bar chart with AI-generated wine bottles. 
To achieve this, \sys integrates GLIGEN~\cite{li2023gligen}, a widely-used Text-to-Image generation model, to ensure that the generated visual elements align with the overall chart structure and the user's intent. 
This automated step accelerates the chart design process while still reflecting the user's design preference.

\stitle{Interactive Refinement.}
In some cases, fully automated methods cannot meet users' needs in a single step.
To address this issue, \sys provides an interface to enable full control over critical design details through interactive refinement.
In this step, users can make fine-grained modifications either by using natural language or by directly interacting with nodes in the chart tree and/or the visual elements.
% Whether it involves adjusting transparency, replacing icons, or refining the layout, this stage enables users to perfect their designs with precision and flexibility, all based on the structure of the chart tree. 
% For additional interaction modes, please refer to Section~\ref{sub:tree_interaction}.

% By balancing AI-driven automation with human-guided refinement and customization, \sys offers a flexible and efficient solution for creating pictorial charts.
% It simplifies the process for users, allowing them to quickly generate visually appealing charts while retaining full control over critical design details.

\subsection{Chart Tree} % : Structuring Chart Elements for Accurate and Consistent Editing
\label{sec:tree}
% \ysy{reviewer:current tree structure is overly simple, and the representation itself lacks support from formative findings. In fact, the current chart tree could be entirely replaced by a two-level configuration panel. While it's acceptable to have a simple instantiation of the chart tree in a prototype system, the paper does not justify the decision to use a tree structure as an intermediate representation. How does it align with users’ mental models? Why is this design more efficient, and how extensible is it for more complex cases? What's more it can afford?}

\begin{figure*}[t!]
    \centering
\includegraphics[width=0.95\linewidth]{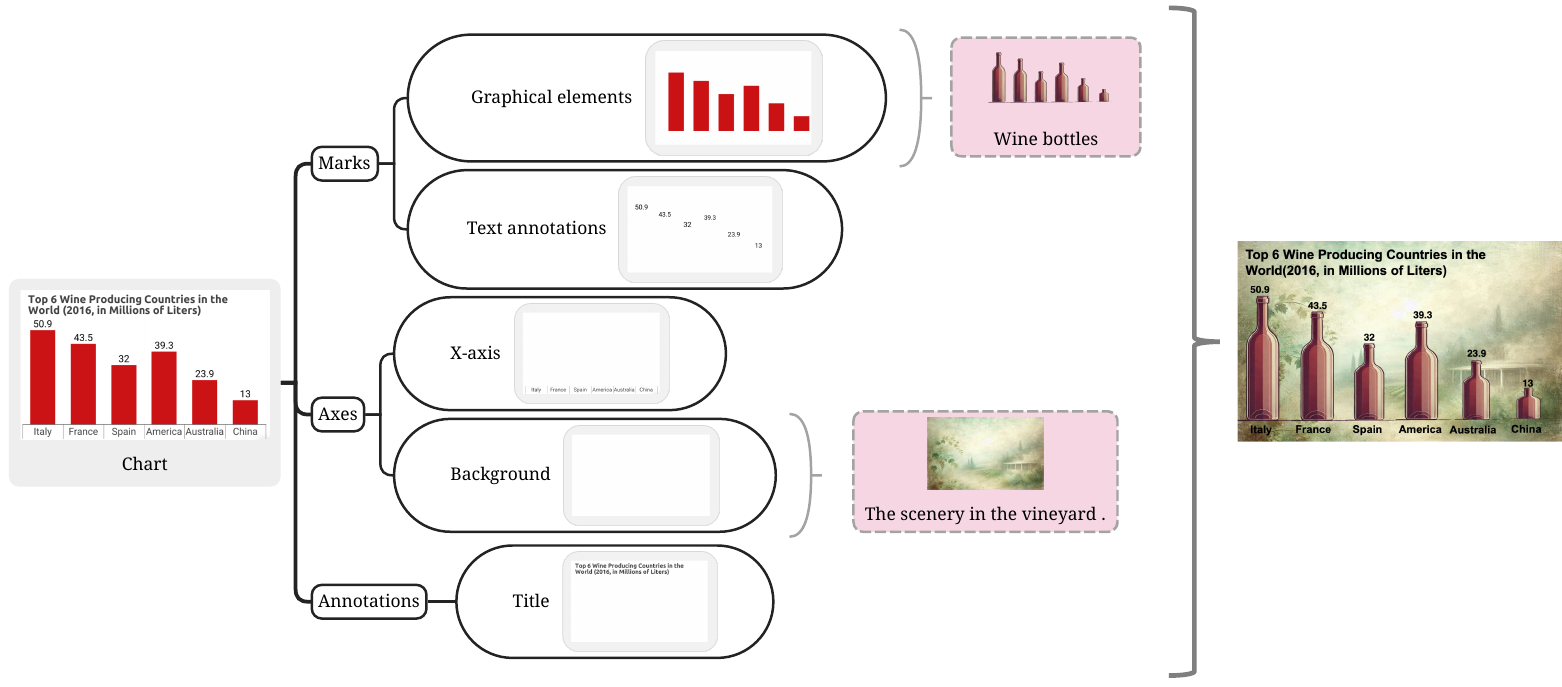}
    \caption{An illustrative example of how the chart tree facilitates the creation of pictorial charts. The first layer (white) represents the decomposition of the chart into its components. The second layer (pink) illustrates the replacement of these components during the auto-generation phase with icons or images imbued with semantic information. Finally, the components are recombined to create the final pictorial chart.}
    \label{fig:chart tree}
\end{figure*} 

The creation of pictorial charts often involves modifying various visual elements. 
However, without a clear framework, these modifications can become complex and inconsistent.
For example, when converting a bar chart into a pictorial one, the bars of the same group and the corresponding legend should be replaced in a consistent manner.
To address this, we introduce the \textbf{chart tree}, a structured framework that organizes visual elements to enable precise and consistent modification.

The chart tree offers two key advantages: (1) it facilitates the automatic generation of pictorial charts by translating high-level user intent into structured, feasible modifications to the tree nodes, and (2) it isolates modifiable visual elements and provides a set of options for manual modification, which provides users full control over the design.

\begin{example}[An Example of Chart Tree]
\textit{\cref{fig:chart tree} illustrates how a simple bar chart can be converted into a pictorial one using the chart tree.
First, the bar chart is decomposed into multiple components, including graphical elements (bars), text annotations, the $X$-axis, the background, and the title.
Next, the bars are replaced with wine bottles, while the background is replaced with a generated scenery.
The modified components are then recombined to create the final pictorial chart.}
\end{example}

\subsubsection{Modifiable Visual Elements}

Guided by the principles of modularity and hierarchical design~\cite{baldwin2000design, saaty1980analytic}, as well as data visualization guidelines~\cite{munzner2014visualization}, we analyzed 1,371 pictorial charts from Pictorial Visualization Dataset~\cite{shi2022supporting} and conducted a comprehensive literature review.
Based on the analysis, we identified a set of key chart components commonly modified during the creation of pictorial charts. 
These components were organized into the chart tree:

\begin{itemize}
    \item \textbf{Marks} are responsible for displaying the primary data elements, such as the bars in bar charts or the lines in line charts.
    In addition to such \textit{graphical elements}, some marks will contain associated \textit{text annotations}, such as the numerical values displayed above the bars, indicating the exact values they represent.
    \item \textbf{Axes} include elements that provide necessary information to understand the values represented by marks.
    This includes the \textit{X-axis}, \textit{Y-axis}, and \textit{backgrounds} with reference lines to aid chart readability.
    \item \textbf{Annotations} provides other information to enhance the readability of charts.
    Here, we considered three types of annotations: \textit{title}, \textit{legend}, and \textit{note} that introduce important insights about the chart. 
\end{itemize}

\begin{figure}[t!]
    \centering        \includegraphics[width=\columnwidth]{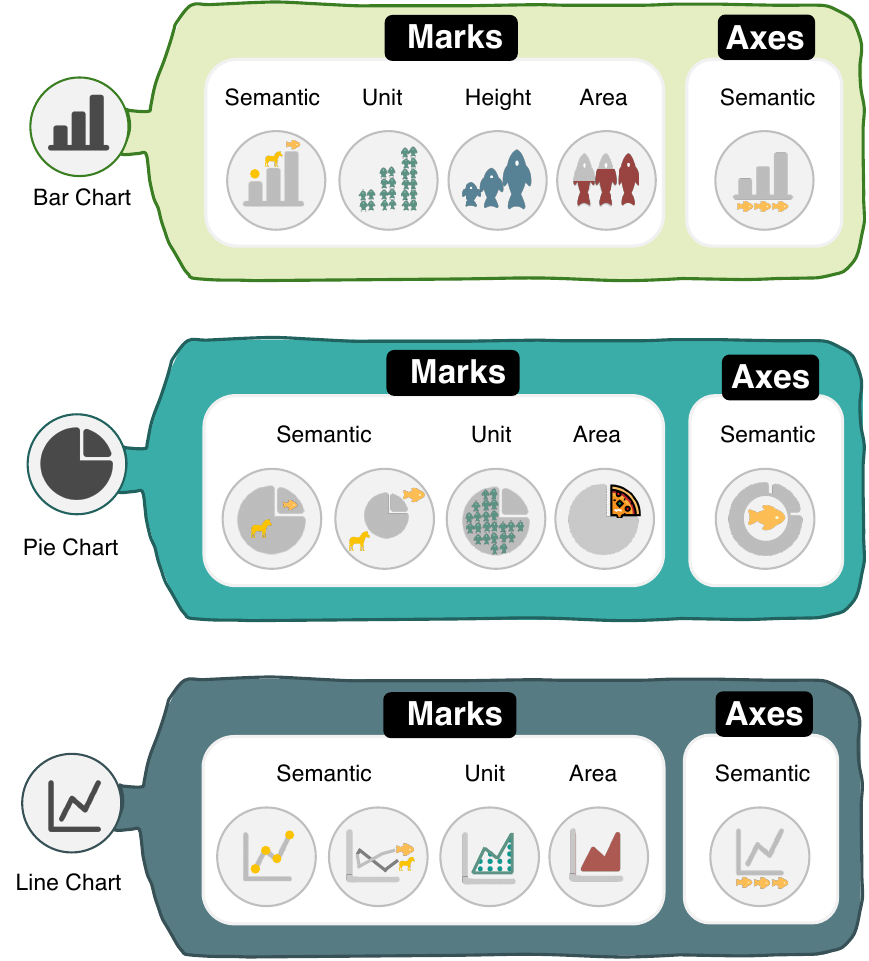}
  \caption{Common design patterns when applying pictorial objects to marks and axes in bar charts, pie charts, and line charts.}
    \label{fig:design}

\end{figure}

\subsubsection{Feasible Modifications}
In addition to identifying modifiable elements, we also identified feasible modifications to these elements.
Beyond common adjustments, such as changing the font size for text or replacing backgrounds with generated images, we emphasized the seamless integration of pictorial elements into marks and axes.
\cref{fig:design} summarizes cases we considered for bar charts, pie charts, and line charts.
Next, we outline specific techniques for integrating pictorial objects into marks and axes.

\etitle{Integrate pictorial objects into marks.}
\label{sec:pic2mark}
This process involves adjusting the pictorial objects to accurately present the values as the original marks.
We summarize four common design patterns\cite{shi2022supporting} to achieve this: semantic, unit, height, and area.

\bi 
\item \textbf{Semantic}.
Semantically relevant objects are widely used to encode categorical data.
For example, in a bar chart illustrating the average numbers of various animals, one may place the icons of corresponding animals on the top of each bar, making the chart easier for readers to interpret at a glance.

\item \textbf{Unit}.
It is also a common practice to use small multiples of pictorial objects to fill the region of chart marks, with the number of units corresponding to the data values or proportions.
For example, in a bar chart showing the number of three different fish species, one may fill each bar with a different number of fish icons, which enhances clarity and helps readers quickly grasp the quantities being represented.

\item \textbf{Height}.
Designers often replace traditional chart marks with stretched pictorial objects. 
For example, in a bar chart showing the number of three different fish species, one may vertically stretch identical fish icons to match the original height of the bars.
However, if the stretch ratio is too extreme, it can result in distorted and visually unappealing representations.

\item \textbf{Area}.
Instead of stretching pictorial objects, designers can maintain uniform object sizes but fill them with proportional colors or cut out pieces to fit specific marks.
This method offers the advantage of maintaining visual uniformity while still conveying quantitative differences.
However, it can be less intuitive for readers, as interpreting color proportion may not be as immediately clear as comparing the heights of objects.
\ei

\etitle{Integrate pictorial objects into axes.}
Since the shapes of the $X$-/$Y$-axes are usually stretched, it is usually not desirable to directly replace the whole axes with pictorial objects.
Therefore, the primary integration method involves replacing the tick labels on the axes with pictorial objects.
For example, text labels that carry semantic information can be replaced with corresponding icons that convey the same meaning.
In addition, we allow users to replace the current background with generated images.

\subsubsection{Modification Modes in Chart Tree}
\label{sub:tree_interaction}
In most cases, there are multiple visual elements that require modification.
Modifying them individually is labor-intensive and may lead to inconsistencies while modifying them all at once limits detailed control.
To address this, the chart tree introduces three modification modes: {\em one-to-one}, {\em one-to-group}, and {\em one-to-all}.
These modes provide varying levels of granularity, enabling users to customize the integration of pictorial objects based on the chart's structure and the data it represents.

\bi 
\item \textbf{One-to-One:} 
This mode allows users to precisely modify a single chart element, such as an individual slide in a pie chart.
It ensures that changes are applied only to the selected element, leaving the rest of the chart untouched, making it ideal for targeted adjustments.
% This mode enables users to modify a single chart element, such as an individual bar in a bar chart.
% Users can select a specific node in the chart tree to apply the desired changes, ensuring that only the selected element is affected while leaving the rest of the chart unchanged.
% This mode is ideal for making precise, targeted adjustments to specific components without affecting others.

\item \textbf{One-to-Group:} 
This mode enables users to modify a subset of visual elements that share characteristics similar to those of the selected element.
For example, in a grouped bar chart, users can apply identical pictorial elements to the first bars in each group, which represent the same category.
Ensure visual consistency across related data categories while preserving flexibility for other parts of the chart.
% This mode allows users to modify a subset of marks that share a common characteristic.
% To achieve this, the user can select a specific group node or a path in the chart tree representing the subset of marks.
% For example, in a grouped bar chart, users can modify the first bars in each group—representing the same category—by applying identical pictorial elements.
% This mode is particularly useful for maintaining visual consistency across related data categories while allowing flexibility for other parts of the chart.

\item \textbf{One-to-All:} 
This mode allows users to apply a single modification to all visual elements of the same type, such as replacing all bars in a bar chart with pictorial elements simultaneously.
By selecting the root or parent node in the chart tree, users can ensure visual consistency across the entire chart, making this mode ideal for achieving uniformity across all elements.
% In this mode, users can apply the same modification to all marks in the chart simultaneously. To achieve this, the user can select the root node or a parent node in the chart tree that represents all marks. For example, in a bar chart, all bars can be replaced with pictorial elements at once, ensuring visual consistency across the entire chart. This mode is ideal for maintaining uniformity across all data points.

\ei

% These interaction modes provide users with flexible and precise editing capabilities, empowering them to integrate pictorial elements in ways that best suit their data and storytelling needs. This level of customization enhances both the visual appeal and the readability of the final chart.

\subsection{Building the Chart Tree Through Chart Decomposition}
\label{sub:decompose}
% We have introduced how to organize chart elements using the chart tree in Section~\ref{sec:tree}.
% In this section, we introduce the process of decomposing the input chart image into individual components and organizing them within the chart tree.

\begin{figure*}[t]
    \centering
\includegraphics[width=.85\linewidth]{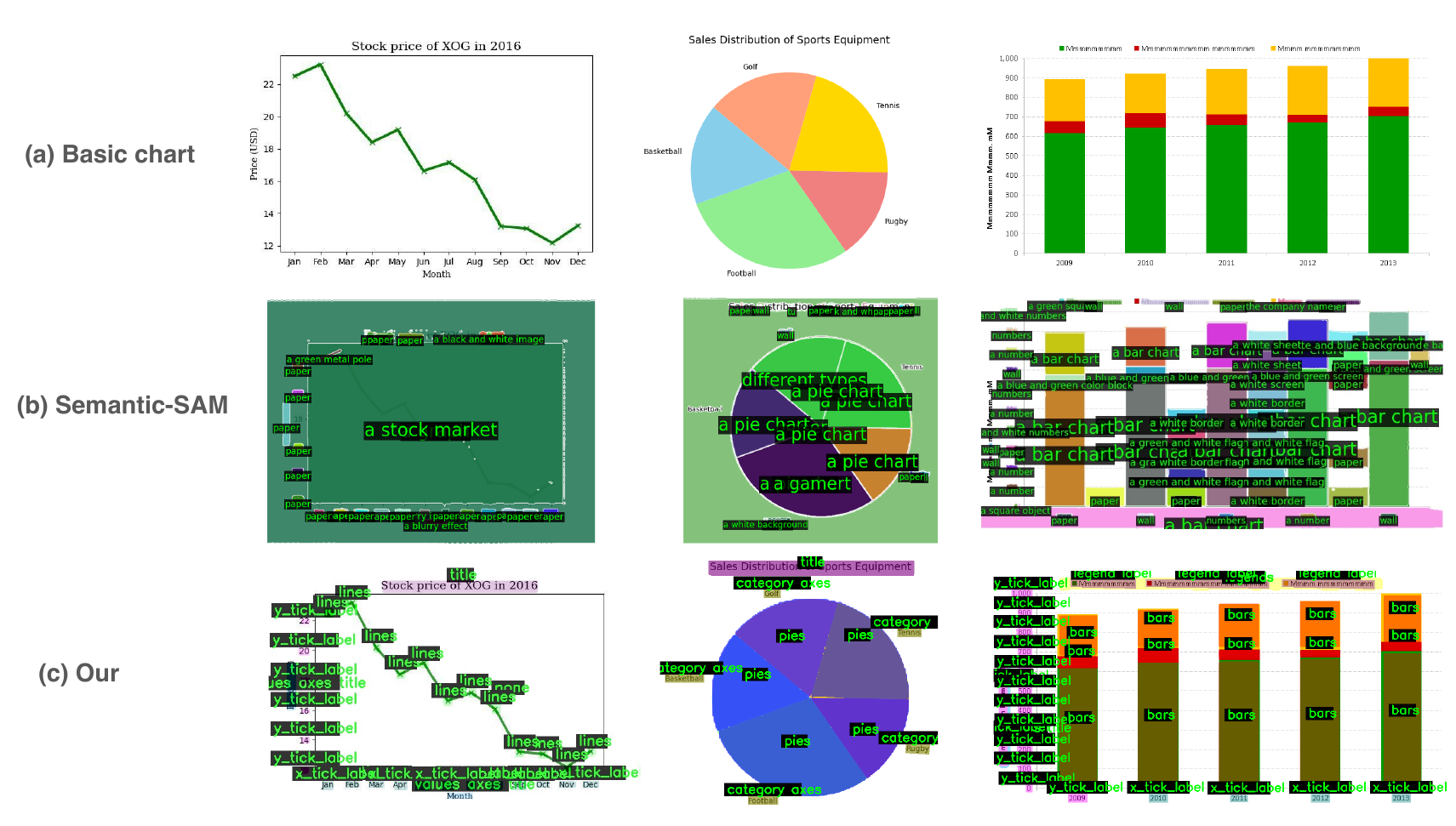}
    \caption{Comparison of chart segmentation results between Semantic-SAM~\cite{li2023semantic} and our method. (a) Shows the original basic charts, (b) Displays the segmentation results from Semantic-SAM, and (c) Illustrates the segmentation results from our approach, highlighting improvements in identifying and labeling key chart elements.}
    \label{fig:segmentation}
\end{figure*}

To take full advantage of the chart tree, the first step is to accurately decompose an input chart image into individual elements and organize them within the chart tree.
A straightforward solution is to leverage existing pre-trained models to segment these elements, such as Semantic-SAM~\cite{li2023semantic}.
However, we found that these models tend to underperform when applied to charts, probably because they are primarily trained on natural images.
\cref{fig:segmentation}(b) highlights three exemplar cases in which Semantic-SAM failed to accurately segment the main marks in the chart and identify their labels.
In addition, the chart segmentation model cannot accurately recognize textual content, which is essential and may require modification during the authoring process.
Therefore, it is necessary to fine-tune a model specifically tailored for the semantic segmentation in charts while integrating Optical Character Recognition (OCR) capabilities~\cite{DBLP:conf/icde/ChaiLFL20}.

\stitle{ChartSS: A Dataset for Chart Semantic Segmentation.}
To improve the segmentation of chart components, we developed ChartSS, a new dataset designed for semantic segmentation in charts. The creation of ChartSS followed a systematic process to ensure diversity, quality, and real-world relevance.

{\textit{Step-1: Dataset Collection.}}
We began by exploring existing datasets published in prior research~\cite{luo2021chartocr,cheng2023chartreader,hassan2023lineex,masson2023chartdetective}.
These datasets provided various charts from real-world scenarios rather than synthesized charts generated from data-generation algorithms.
To further increase the diversity of the dataset, we also gathered a substantial amount of annotated chart data from the online repositories Roboflow Universe~\cite{roboflow_universe}.

{\textit{Step-2: Dataset Refinement.}}
To maintain quality and usability, we screened the collected data to exclude overly complex images that could hinder segmentation tasks. This curation process resulted in a final dataset of 59,693 images, comprising a balanced mix of 31,427 bar charts, 9,946 line charts, and 18,320 pie charts.

{\textit{Step-3: Dataset Splitting.}}
The curated dataset was then partitioned into training (70\%), validation (20\%), and testing (10\%) subsets to support model development, fine-tuning, and evaluation. This split ensures a robust framework for assessing segmentation models while minimizing overfitting risks.

\stitle{OCR-Assisted Mask2Former.}
To fine-tune a chart segmentation model on the ChartSS dataset, we began by evaluating several state-of-the-art semantic segmentation models to determine the most effective approach. These included Mask2Former~\cite{cheng2021maskformer} (a Transformer-based model), DeepLabV3+\cite{Chen_2018} (a classical convolutional model), and YOLOv8-Seg\cite{yolov8} (from the YOLO series). Among these, Mask2Former demonstrated superior performance in capturing long-range dependencies and fine details, making it the best choice for chart segmentation. Please refer to Section~\ref{exp1} for experimental results.

However, while Mask2Former excels in segmenting visual elements, it lacks the ability to recognize textual content, which is essential for further editing. To address this limitation, we integrated Mask2Former with CnOCR~\cite{cnocr_2023}, enabling accurate extraction of text from various types of charts.

To achieve this, we first determine if the detected text overlaps with a segmented region. If it does, the text is directly assigned to the corresponding component. For text outside any segmented region, heuristic chart rules are applied. For example, text located beneath the $X$-axis is classified as an $X$-tick label. This approach ensures that all textual elements, including those missed by the Mask2Former model, are accurately categorized.

As illustrated in \cref{fig:segmentation}(c), our method effectively segments chart elements such as bars, lines, slices, $X$-axes, and legends while accurately extracting text, ensuring comprehensive chart understanding and editability.

% To fine-tune a chart segmentation model on the ChartSS dataset, we first evaluated multiple state-of-the-art semantic segmentation models, including Mask2Former~\cite{cheng2021maskformer} (a transformer-based model), DeepLabV3+ (a classical convolutional model)~\cite{Chen_2018}, and YOLOv8-Seg~\cite{yolov8} (from the YOLO series) to identify the best model for chart segmentation.
% The results showed that Mask2Former produces better results (see Sec.~\ref{exp1} for more details) due to its ability to capture long-range dependencies and finer details.
% However, chart segmentation models are not capable of recognizing textual content for further editing.
% To address this, we combine it with CnOCR~\cite{cnocr_2023}, to accurately extract text from various types of charts.
% \cref{fig:segmentation}(c) shows the segmentation and OCR results of our method, which accurately identifies various elements such as bars, lines, slices, $X$-axis, and legends.

% Both the open-sourced dataset and the segmentation model checkpoints are available at \url{https://github.com/Chart-Editor/ChartEditor.git}.
% By combining Mask2Former with CnOCR, we introduce the OCR-Assisted Mask2Former, which leverages both graphical and textual segmentation, improving overall performance.
% For textual components, we employ the OCR technique, which has proven effective in locating and extracting text from charts.
% Specifically, we use CnOCR~\cite{cnocr_2023} to accurately extract text from various types of charts in our system.

\stitle{Chart Tree Construction based on Segmentation and OCR.}
The segmentation results are systematically organized into the chart tree as connections between segmented elements and the chart tree are straightforward due to predefined segmentation labels.

% The segmentation and OCR results are systematically organized into the chart tree. While connections between segmented elements and the chart tree are straightforward due to predefined segmentation labels, linking OCR-detected text is more challenging, as text can appear in various locations such as titles, axis labels, and legends.

% To address this, we first determine if the detected text overlaps with a segmented region. If it does, the text is directly assigned to the corresponding component. For text outside any segmented region, heuristic chart rules are applied. For example, text located beneath the $X$-axis is classified as an $X$-tick label. This approach ensures that all textual elements, including those missed by the Mask2Former model, are accurately categorized and integrated into the chart tree.

In addition, we establish links between marks and legends by analyzing their color and texture. Marks and legends with matching appearances are grouped together, allowing for consistent modifications and ensuring visual uniformity across the chart.
% The segmentation and OCR results are then organized within the chart tree.
% The connections between the segmentation results and the chart tree can be easily established based on the segmentation labels.
% However, linking OCR results to the chart tree is more complex because text can appear in various areas, such as titles, axis labels, and legends.
% To correctly establish these connections, we first check whether the detected text is located within a segmented region.
% If it is, the text is assigned to the corresponding component.
% If the text lies outside any segmented region, we apply heuristic chart rules. 
% For example, text located directly beneath the x-axis is classified as an $X$-tick label. 
% This ensures that any text not detected by the Mark2Former model is accurately categorized and integrated into the chart tree.
% In addition, we establish connections between marks and legends based on their color and texture.
% Marks and legends with the same appearance are grouped together and will receive the same modifications to ensure visual consistency.

% \subsection{Automatic Pictorial Chart Generation with Idea Prompting}
\subsection{Automatic Generation}
\label{sub:autosys}
Once the chart tree is constructed, users can express their design intent in natural language.
Since user descriptions are often brief and vague, idea prompting is used to translate this high-level intent into feasible modifications to the chart tree~\cite{DBLP:journals/pvldb/LiLCLT24,DBLP:journals/corr/abs-2406-07815,DBLP:journals/corr/abs-2408-05109}. 
These modifications are then applied to generate the initial pictorial chart.

\subsubsection{Idea Prompting}
\begin{figure}[t!]
\centering
\includegraphics[width=1\linewidth]{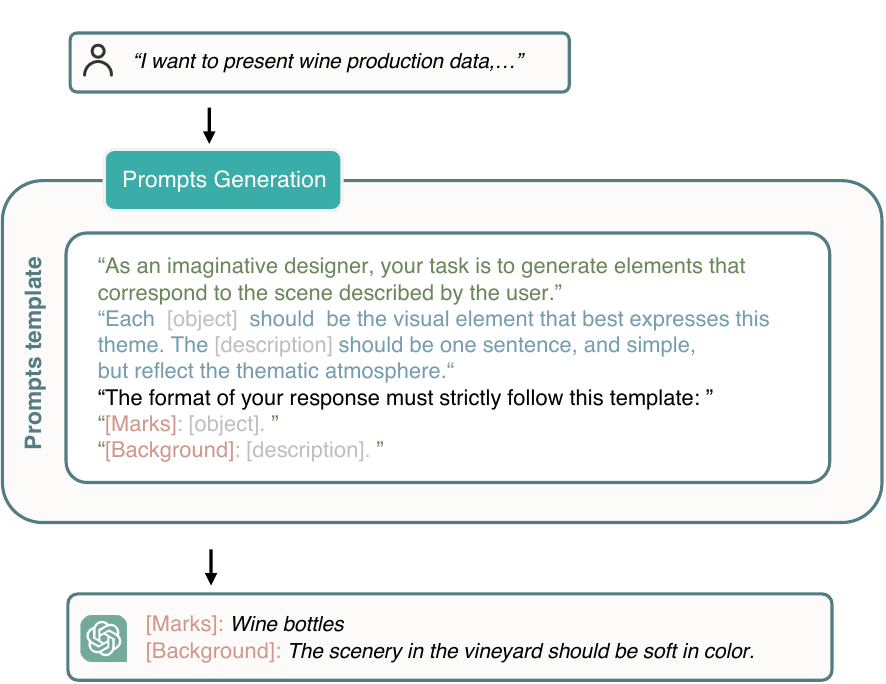}
\caption{Idea prompting process: Transforming fuzzy descriptions into specific modifications using a defined prompt template.}
\label{fig:idea}
\end{figure}

% In the automatic generation stage, we directly generate the pictorial charts based on the extracted chart tree and user intent.
% However, user input is usually brief and fuzzy.
% Idea prompting is designed to convert high-level user intent into actionable modifications to the chart tree.
When users begin to create a pictorial chart, they usually lack a clear design plan but only provide some brief or vague design intent.
For instance, a user might say, ``I want to present wine production data", which by itself is insufficient for directly generating a pictorial chart.
Inspired by previous research~\cite{hou2024c2ideas,wang2022towards}, we use a defined prompt template to translate high-level user intent into feasible modifications to the chart tree.
An illustrative example is provided in \cref{fig:idea}, where the vague intent is translated into specific editing actions, including replacing chart marks with pictorial objects of wine bottles, and adjusting the background to feature a vineyard scene in soft colors.

% This serves as a bridge between the user and the visual element generation. We have designed a prompt template to facilitate the generation of editing actions, consisting of three components, as shown in Figure~\ref{fig:idea}.
% For example, the aforementioned fuzzy intent can be translated into specific editing actions, including objects such as ``wine bottles" and background attributes like ``vineyard" and ``soft color''.

\subsubsection{Automatically Applying Modification}
Once the desired modifications are determined through idea prompting, the next step is to apply them to the relevant nodes in the chart tree.
% Simple modifications, such as changing font colors or adjusting positions, can be applied immediately.
However, substantial modifications, like replacing the main mark or background, require multiple steps.
These typically involve (1) generating pictorial objects that align with the user's intent, and (2) integrating them into the chart.

\begin{itemize}
\item \textbf{Generating Pictorial Objects}.
The first step is to generate a corresponding pictorial object that reflects the design intent.
This process requires two inputs.
The first one is the outlines of the visual elements associated with the selected node, which serve as a mask during the diffusion process.
The second one is the description of objects, which is generated during idea prompting.
The resulting pictorial objects can be used directly without complex operations, significantly reducing efforts in searching and post-processing.
% The second input corresponds to the editing action associated with this node, provided by the idea prompting process.

\begin{figure*}[t!]
    \centering
\includegraphics[width=1\textwidth]{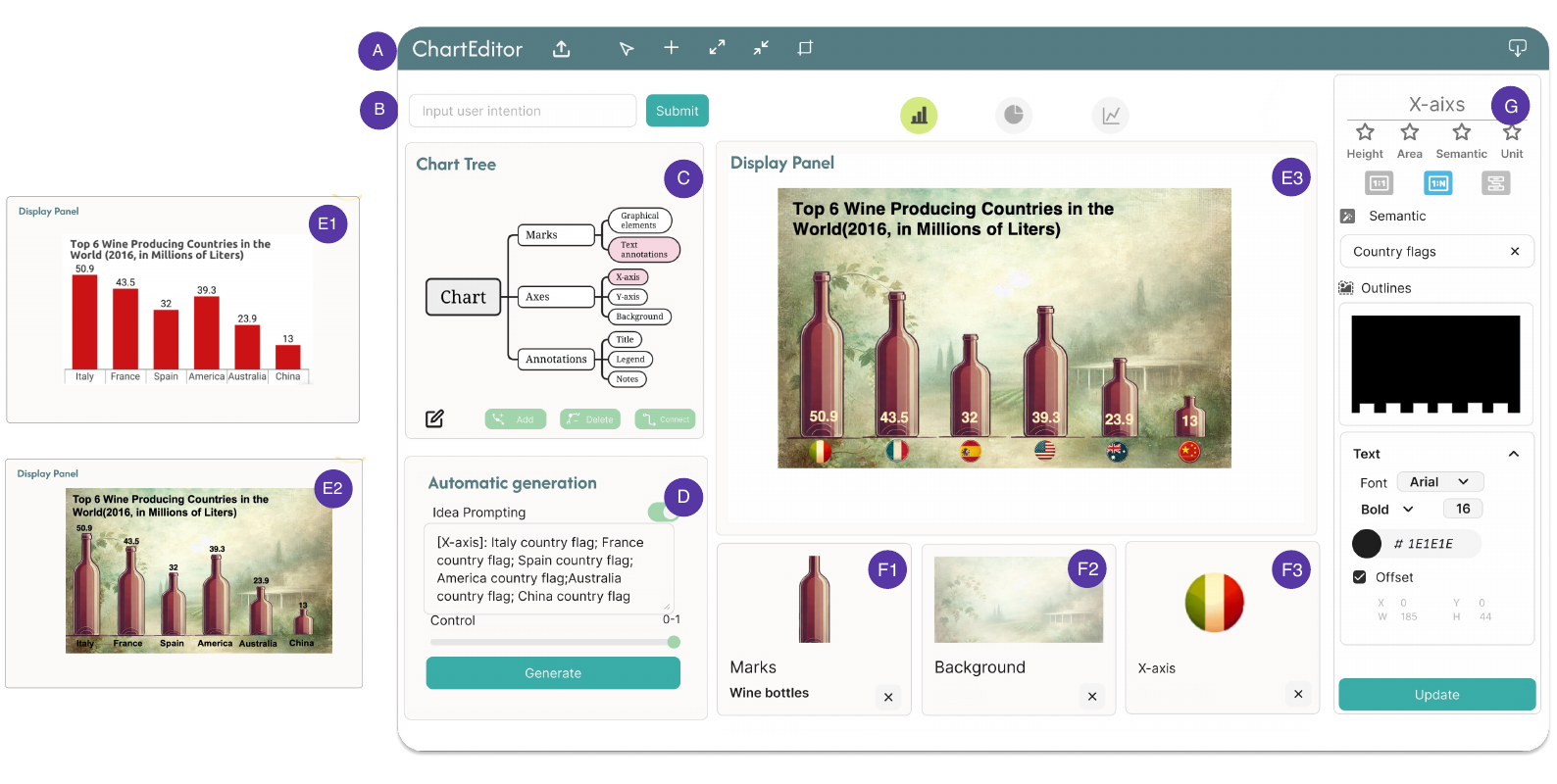}
    \caption{The interface of \sys features a navigation bar (A) for uploading, downloading, and accessing tools. After uploading, the chart appears in the initial display panel (E1). The chart’s components are shown in a chart tree (C), where they can be edited. Users can input their intent and click the submit button (B), prompting suggestions in the automatic generation panel (D). Changes are applied and reflected in the modified display panel (E2). Additional customizations, like flags, generate new edits in the chart tree (F1, F2, F3). Detailed adjustments can be made in the node edit panel (G). The final version is shown in the final display panel (E3).}
    \label{fig:UI}
    
\end{figure*}

\item \textbf{Integrate Pictorial Objects into the Chart}.
When we use the pictorial object as a background, we can simply replace the original background with this object.
However, if the object serves as the main mark, a more detailed process is required.
First, we identify the outlines of the main marks to determine placement areas for the pictorial objects.
We then integrate them into the corresponding positions using the four methods defined in Section~\ref{sec:pic2mark}.
For categorical data, \sys adopts \textit{semantic} by default.
For numerical data, \sys applies different methods depending on the chart type: \textit{height} for bar charts, \textit{area} for pie charts, and \text{semantic} for line charts.
These are the most common choices based on our analysis of the existing pictorial charts.
\end{itemize}

\subsection{Interactive Refinement with the Chart Tree}
\label{sub:intersys}

While \sys is capable of automatically generating high-quality pictorial charts, the results do not always align perfectly with user expectations.
Many users prefer to customize the charts further to communicate their design intent more effectively.
To accommodate this, we have introduced an interactive interface (\cref{fig:UI}) that allows fine-grained adjustments based on the automatically generated charts.
Initially, users input their chart and design intent through this interface.
Then, they can examine the automatically applied modifications and the corresponding generated results.
Targeted modifications can be made by selecting either the nodes of the chart tree or the visual components of the charts.
More details of this interface will be introduced through a usage scenario in Section~\ref{sec:usage}.
\section{Usage Scenario}
\label{sec:usage}

Imagine a marketing student, Salin, who wants to transform a basic chart of wine production data into a visually engaging pictorial chart for her presentation to impress her teachers. With our \sys, Salin can effortlessly convert the basic chart into its pictorial version, as illustrated in \cref{fig:UI}.

Initially, Salin uploads the basic chart to the navigation bar~(A), which is simultaneously displayed in the display panel~(E1).
\sys automatically decomposes the chart into multiple visual elements, including marks, annotations, and axes.
The corresponding nodes are now modifiable in the chart tree panel~(C).
% She finds that chart tree (C) will highlight the components detected in the input chart, and the corresponding nodes in the tree are now editable. 
% In cases where the chart contains unconventional components, users are provided with the option to add, delete, or connect nodes within the chart tree, allowing for the establishment of relationships between nodes as necessary.

Next, Salin inputs her intent, ``\textit{I want to make this wine production chart more memorable}'', and clicks the ``Submit'' button (B).
The automatic generation panel (D) then displays a refined suggestion based on the idea-prompting process.
She is satisfied with it and clicks on the ``Generate'' button.
Two modifications are then applied to the chart tree, including replacing marks with wine bottles and adding a background image of the vineyard.
The panel (E2) shows the resulting pictorial chart.
Salin is impressed by the outcome as it is visually appealing and aligns well with the original data.
Therefore, Salin decides to use this as a starting point and customize it further to enhance the chart's quality.
% She proceeds to the automatic generation panel (D), where she finds a refined suggestion provided by the idea prompting.
% Upon clicking the ``Generate'' button, the modifications applied to the chart tree (F1, F2) and the resulting pictorial chart (E2) are displayed.
% Salin is impressed by the outcome as it is visually appealing and aligns well with the original data.

First, Salin notices that the text annotations above the bars become less readable after adding the background image.
She decides to move the annotations inside the bars, switch the font color to beige color, and increase the font size to improve readability.
To achieve these, she clicks the node ``Marks - Text Annotations'' in the chart tree, and the feasible modifications are displayed in the node edit panel (G).
Since she needs to adjust the y-positions, font colors, and font sizes to the same values for all annotations, she opts for the ``one-to-all'' mode \includegraphics[width=0.4cm]{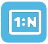} to modify all annotations via a single click.
% The corresponding region being edited is also displayed in the middle of the node edit panel.
% If she wants to modify a specific annotation, such as highlighting the largest values in red, she can switch to ``one-to-one'' mode \includegraphics[width=0.4cm]{figure/icon/one-to-one.pdf}.

Next, Salin wants to replace the country names with their respective flags to enhance visual recognition and appeal.
Therefore, she inputs the ``country flag'' in the automatic generation panel (D).
Upon clicking ``Generate'', a new modification (F3) is automatically generated.
She clicks it to reveal the details of this modification.
This modification involves generating semantically relevant pictorial icons (national flags) for each country and integrating them into the node ``$X$-axis'' in a ``semantic'' manner.
% Salin is satisfied that \sys correctly understands her intent and applies correct modifications to the chart.

% She observes that a new sticker has been applied to the selected region, and it appears correctly in the display panel (E2). She proceeds to modify the rest of the x-axis in a similar manner, finally pressing the "Generate" button again to finalize the chart.

Finally, she adjusts the transparency of the background image to enhance the chart's visual appeal. This is done by selecting the background within the chart panel and modifying its transparency using the node edit panel.
% Finally, she directly adjusts the transparency of the background image to achieve a more visually appealing result.
% She does this by clicking on the background within the chart panel and modifying its transparency in the node edit panel.

The completed pictorial chart is displayed in the display panel (E3). Pleased with the result, Salin clicks the download icon \includegraphics[width=0.4cm]{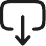} {(A)} to save the chart. She appreciates how \sys simplifies the process of creating pictorial charts.

\section{Quantitative Evaluation}
\label{sec:quan}

% Given that system performance heavily depends on the accuracy of each module, we assess the significance of chart decomposition.
% Given that system performance largely relies on the quality of the chart tree, we conducted a quantitative evaluation focusing on the performance of chart decomposition.
% In addition, we conducted a user study to evaluate the usability of \sys in authoring pictorial charts.
% \subsection{Quantitative evaluation}
Given that system performance largely relies on the quality of the chart decomposition, we conducted a quantitative evaluation to assess the performance of our chart segmentation method and the usefulness of the resulting decomposition.
% \subsection{Identify the Best Chart Segmentation Model}
\subsection{Performance of Chart Decomposition Method}
\label{exp1}
% As previously mentioned, we fine-tuned multiple models using the ChartSS dataset. Below, we outline the comparison process.
\stitle{Baselines}.
We evaluated three state-of-the-art semantic segmentation models for chart component segmentation:
\begin{itemize}
    \item Mask2Former~\cite{cheng2021maskformer}, a Transformer-based model.
\item DeepLabV3+~\cite{Chen_2018}, a classical convolutional model.
\item YOLOv8-Seg~\cite{yolov8}, a segmentation model from the YOLO series.
\end{itemize}

For Mask2Former and DeepLabV3+, we tested two widely used backbone architectures: ResNet-50 and ResNet-101.
For YOLOv8-Seg, the default backbone CSPNet was used.

\stitle{Training Details}.
All models were initialized with pre-trained weights and fine-tuned on our ChartSS dataset. We used the AdamW optimizer with a learning rate of 0.0001 and a batch size of 8. Each model was trained for 12 epochs on an NVIDIA A800 GPU, ensuring consistent conditions across experiments.

This setup allowed us to fairly compare the performance of different models and backbone configurations on the ChartSS dataset.

% We compared three state-of-the-art semantic segmentation models: Mask2Former\cite{cheng2021maskformer} (a Transformer-based model), DeepLabV3+ (a classical convolutional model)\cite{Chen_2018}, and YOLOv8-Seg\cite{yolov8} (YOLO series model).
% For both Mask2Former and DeepLabV3+, we tested two widely used backbone architectures: ResNet-50 and ResNet-101.
% For YOLOv8-Seg, we used the default backbone CSPNet.
% All the models were initialized with pre-trained weights and fine-tuned on our ChartSS dataset.
% We utilize the AdamW optimizer with a learning rate of 0.0001 and a batch size of 8, and the training was performed for 12 epochs on an NVIDIA A800 GPU. 

% \stitle{Implementation details.}
 % In this study, we only report results using the R101 backbone.  Both models were initialized with pre-trained weights and fine-tuned on our dataset. We utilize the AdamW optimizer with a learning rate of 0.0001, a batch size of 8, and the training was performed for 12 epochs on NVIDIA A800 GPUs. 

\stitle{Metrics}.
In line with standard practices in semantic segmentation, we used mean Intersection over Union (mIoU)~\cite{DBLP:journals/corr/RonnebergerFB15} as the evaluation metric. mIoU quantifies the overlap between predicted segmentation masks and ground truth masks, averaged across all chart components.
% calculates the mean IoU across all chart components.

\begin{table}[t!]
\centering
\caption{Comparison of segmentation performance across DeepLabV3+, YOLOv8-Seg and Mask2Former.}
\label{tab:segmentation_comparison}
\resizebox{0.9\columnwidth}{!}{ % 调整宽度以适应内容
\begin{tabular}{l  c c c}
\toprule
\textbf{Model} & \textbf{Backbone} & \multicolumn{2}{c}{\textbf{mIoU (\%)}} \\ 
\cmidrule(lr){3-4} % 在mIoU列下划分子列
 &  & \textbf{Pre-trained} & \textbf{Fine-tuned} \\
\midrule
\multirow{2}{*}{DeepLabV3+} 
& ResNet-50 & 3.82 & 52.86 \\
& ResNet-101  & 5.16 & 54.16 \\
\midrule
YOLOv8-Seg & CSPDarkNet     & - & 54.30 \\
\midrule
\multirow{2}{*}{Mask2Former (ours)} 
& ResNet-50 & 5.57 & 76.5 \\
& ResNet-101 & 7.05 & \textbf{78.90} \\
\bottomrule
\end{tabular}
}
\label{tab:segmentation_comparison}
\end{table}

\stitle{Results}.
As shown in Table \ref{tab:segmentation_comparison}, all models demonstrate a substantial improvement in mIoU after fine-tuning on our ChartSS dataset.
This emphasizes the disparity between chart images and pretraining images, underscoring the critical need for our domain-specific dataset collection.
%, underscoring the gap between the chart images and their pretraining images.
% Notably, YOLOv8-Seg does not present a pretrained metric because its COCO-based categories diverge entirely from our chart classes, rendering initial performance negligible.
Meanwhile, deeper backbones consistently surpass shallower ones: ResNet-101 outperforms ResNet-50 in capturing the nuanced patterns within charts.
Among all models, Mask2Former achieves the highest accuracy at 78.90\%, likely due to its transformer-based architecture, which excels at capturing global context compared to purely convolutional or YOLO-style approaches. 
Therefore, we selected Mask2Former with a ResNet-101 backbone for our method, as it proved to be the most effective choice for our chart segmentation task.

\begin{table}[t!]
\centering
\caption{Comparison of style transfer methods with and without Decomposition support. Metrics include LPIPS, Color Accuracy, and User Rating (1-5).}
\label{tab:style_transfer_comparison}
% \resizebox{0.8\columnwidth}{!}{ % 调整为列宽度

\begin{tabular}{ l c c c }
\toprule
\textbf{Method}           
& \textbf{LPIPS} & \textbf{SA (\%)} & \textbf{User Rating (1-5)} \\ \midrule
GLIGEN & 0.55 & 65   & 3.2   \\

GLIGEN+Decomposition & \textbf{0.35} 
&\textbf{85}  & \textbf{4.5}  \\ \bottomrule

\end{tabular}
\end{table}

\begin{figure*}[t!] % H forces the figure to be placed exactly here
    \centering
    \includegraphics[width=\textwidth]{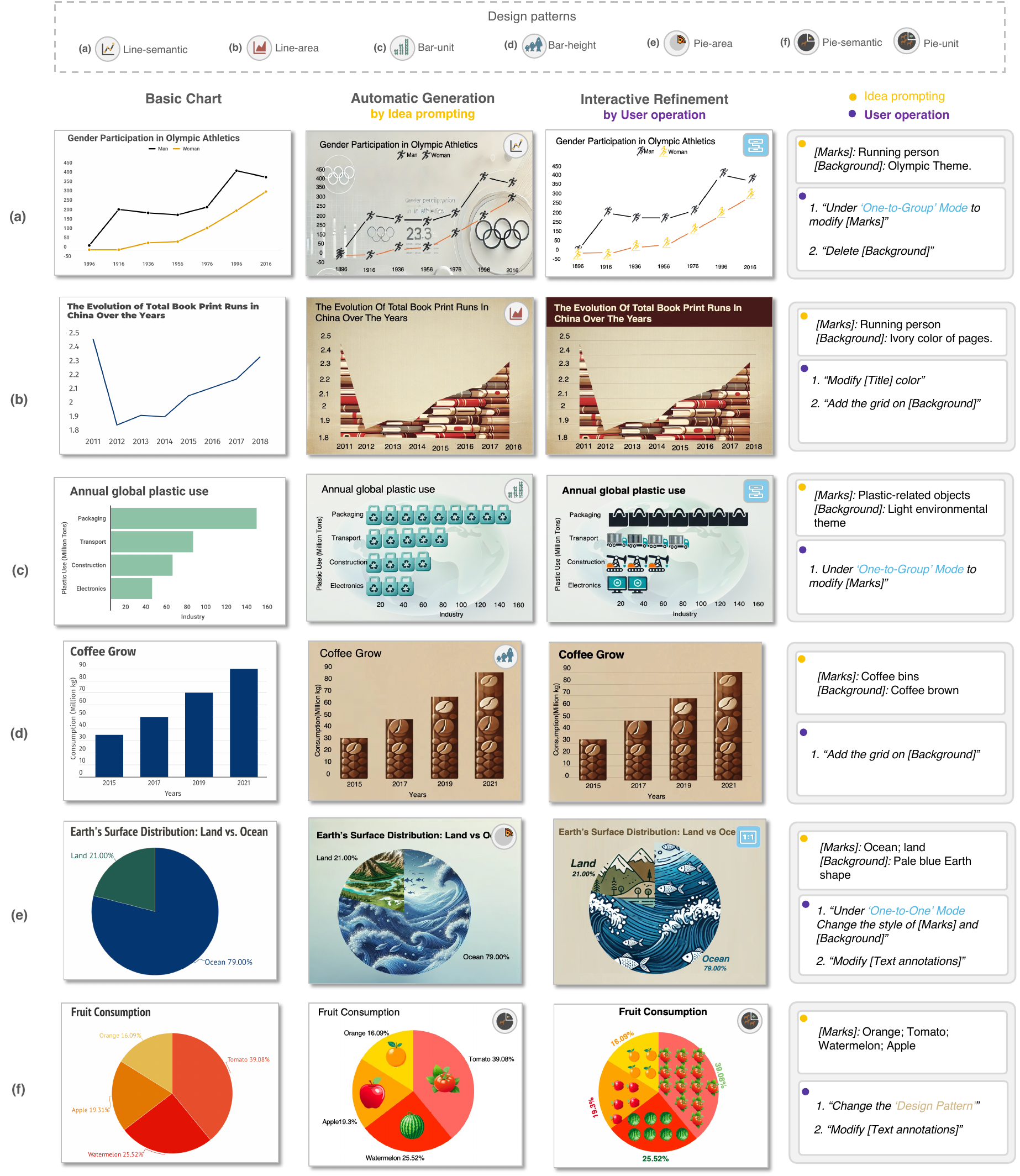} \\ % First row
    \caption{Example pictorial charts created with \sys in User Study. (a) shows a line-semantic chart theme; (b) presents a line-area chart theme; (c) illustrates a bar-unit chart theme; (d) shows a bar-height chart theme; (e) shows a pie-area chart theme; (f) presents a pie-semantic and pie-unit chart theme.}
    \label{fig:gallery}
\end{figure*}

% \subsection{Comparison on Chart Decomposition}
% \subsection{Evaluate the Usefulness of Chart Decomposition}
\subsection{Usefulness of the Chart Decomposition Results}
\stitle{Experimental Setup}.
% To evaluate the usefulness of chart decomposition in the generation of pictorial charts. We compared applying GLIGEN\cite{li2023gligen} after chart decomposition with those generated by GLIGEN directly, both using descriptions generated through Idea Prompting.
We evaluate the usefulness of chart decomposition in generating pictorial charts by comparing charts produced with and without chart decomposition.
We randomly selected nine charts from our proposed dataset ChartSS as input, including three bar charts, three line charts, and three pie charts.
Each chart was assigned a distinct target style for transformation.
The baseline method directly applies GLIGEN~\cite{li2023gligen} based on the descriptions generated through Idea Prompting, and our method uses both descriptions and chart decomposition results when applying GLIGEN.
% This comparison helps assess whether chart decomposition contributes to the quality of the generated charts.
% \stitle{Data.}

\stitle{Metrics}.
We used the following three metrics to assess the quality of the generated pictorial charts:
\begin{itemize}
\item Learned Perceptual Image Patch Similarity (LPIPS): Measures perceptual differences between images. Lower values indicate better perceptual similarity.
\item Style Accuracy (SA): Assesses the consistency of artistic or stylistic features between images. Higher values signify closer stylistic alignment.
\item Mean Opinion Score (MOS): Subjective ratings were collected from eight participants who evaluated the generated charts on a 1–5 scale across four dimensions: data preservation, clarity, aesthetics, and overall satisfaction. The average score was used as the final MOS.
\end{itemize}

\stitle{Results}.
% Table~\ref{tab:style_transfer_comparison} demonstrates that the approach with chart decomposition consistently outperformed the direct GLIGEN, achieving lower LPIPS scores, indicating a more significant perceptual difference and better reconstruction of the original chart features.
Table~\ref{tab:style_transfer_comparison} demonstrates that GLIGEN performs better with chart decomposition across all the metrics.
The lower LPIPS score indicates that chart decomposition helps maintain the original chart structure and produce visually similar results.
Our method also achieves a higher SA score, indicating a more consistent style.
% While both methods showed relatively low SA scores due to a focus on style transformation rather than preserving the original chart design, this highlights a key area for future improvement.
Notably, the approach with chart decomposition achieved higher MOS ratings, reflecting greater alignment with human visual preferences.
These results demonstrate the effectiveness of the chart decomposition module in improving the quality of pictorial chart generation.

\section{User Study}
\label{sec:userstudy}
% We conducted a comparative experiment to evaluate the effectiveness of \sys. In the experiment, we chose  DataQuilt\cite{zhang2020dataquilt} MetaGlyph~\etal\cite{ying2022metaglyph} as baselines.

\subsection{Study Design}
To thoroughly evaluate the usability of \sys, we conducted a user study and benchmarked it against previously discussed systems, DataQuilt\cite{zhang2020dataquilt} and MetaGlyph\cite{ying2022metaglyph}.
In addition, we curated a gallery showcasing participants’ outputs generated during the process.

\stitle{Participants}. Through the school’s mailing list, we recruited and screened 18 participants with entry-level design experience. % and all were required to be on-site via random sampling
The participants ranged in age from 18 to 41 and had diverse educational backgrounds and fields of study.
The group included one administrative staff member (S1), seven undergraduate students (S2–S8), seven postgraduate students (S9–S15), and three assistant professors (S16–S18).
% All of them share a need to author data into charts.
All of them share a need to create pictorial charts that engage their readers.
Based on their answers on a five-point Likert scale (where one strongly disagreed and five strongly agreed), participants claimed that they were willing to employ AI-assisted tools to create charts (M = 3.94, SD = 0.87). 

\stitle{Methods}.
% This study aims to evaluate the workload, effectiveness, and expressiveness of the \sys system, comparing it with two alternative systems: MetaGlyph \cite{ying2022metaglyph} and DataQuilts \cite{zhang2020dataquilt}. 
This study aims to evaluate the workload, effectiveness, and expressiveness across different systems.
To achieve this, we employed the NASA Task Load Index (NASA-TLX) and a five-point system evaluation scale for quantitative feedback, while think-aloud sessions and semi-structured interviews were conducted for qualitative data collection. 
All interview recordings were transcribed using the iflyrec platform and analyzed thematically to identify common patterns and insights from participants’ experiences with the three systems.

% All interview recordings were transcribed into text using the iflyrec platform for further analysis.

% The study aimed to evaluate the system’s workload, effectiveness, and expressiveness by comparing it against two alternative systems—MetaGlyph and DataQuilts.
% Participants were assigned a chart from a provided basic chart library and use \sys and the other systems to generate a pictorial chart in random order.
% Their experiences and perceptions were assessed using two standardized questionnaires, the NASA Task Load Index (NASA-TLX)~\cite{hart2006nasa} and a five-point system evaluation scale.
% Qualitative data were also collected during think-aloud sessions and semi-structured interviews.
% The audio recordings from these sessions were transcribed into dialogue-format scripts using the {\em iflyrec} platform for further analysis.

\stitle{Study Procedure}.
After a brief tutorial and a warm-up task, participants were randomly assigned a chart theme and asked to create charts using \sys, MetaGlyph, and DataQuilts in a random order.
This randomization of order was specifically implemented to eliminate biases caused by the task sequence.
During the task, participants were instructed to think aloud, verbalizing their decision-making processes to provide real-time insights.
Upon completing the tasks, participants filled out the NASA-TLX and the system evaluation scale, followed by semi-structured interviews to gather in-depth qualitative feedback.
The study was conducted in a UX research room equipped with a one-way glass window, enabling researchers to observe and document participant interactions unobtrusively.
\subsection{{Pictorial Chart Gallery}}
We also present the results of \sys generated by participants under various themes during the experiment.
For each theme, we selected the most visually compelling outcome, as shown in \cref{fig:gallery}.
The first column presents the basic chart, while the second column displays the results of automatic generation.
The top right corner indicates different recommended design patterns for each chart type, which highlights the versatility of our method.
The third column illustrates the outcomes of interactive generation and notes the corresponding operation mode, while the last column details the process.

\subsection{Result Analysis}
\subsubsection{Quantitative Results}

We compared the NASA Task Load Index and the System Evaluation Five-Point Scale of three systems horizontally by radar chart. 

% \begin{figure}
%     \centering
%     \includegraphics[width=1\linewidth]{figure/radar-compare.png}
%     \caption{Radar chart presenting the comparative results of user study data for different chart authoring tools. \textbf{The data has been normalized, the purpose of which is to amplify the minor differences and make the contrast more obvious.}}
%     \label{fig:radar}
% \end{figure}

% \begin{figure}
%     \centering
%     \includegraphics[width=1\linewidth]{figure/radar-.png}
%     \caption{Radar chart presenting the comparative results of user study data for different chart authoring tools. \textbf{The data has been normalized, the purpose of which is to amplify the minor differences and make the contrast more obvious.}}
%     \label{fig:radar}
% \end{figure}

% \begin{figure}
%     \centering
%     \includegraphics[width=1\linewidth]{radar-2.png}
%     \caption{Radar chart presenting the comparative results of user study data for different chart authoring tools. \textbf{The data has been normalized, the purpose of which is to amplify the minor differences and make the contrast more obvious.}}
%     \label{fig:radar}
% \end{figure}

\begin{figure}
    \centering
    \includegraphics[width=1\linewidth]{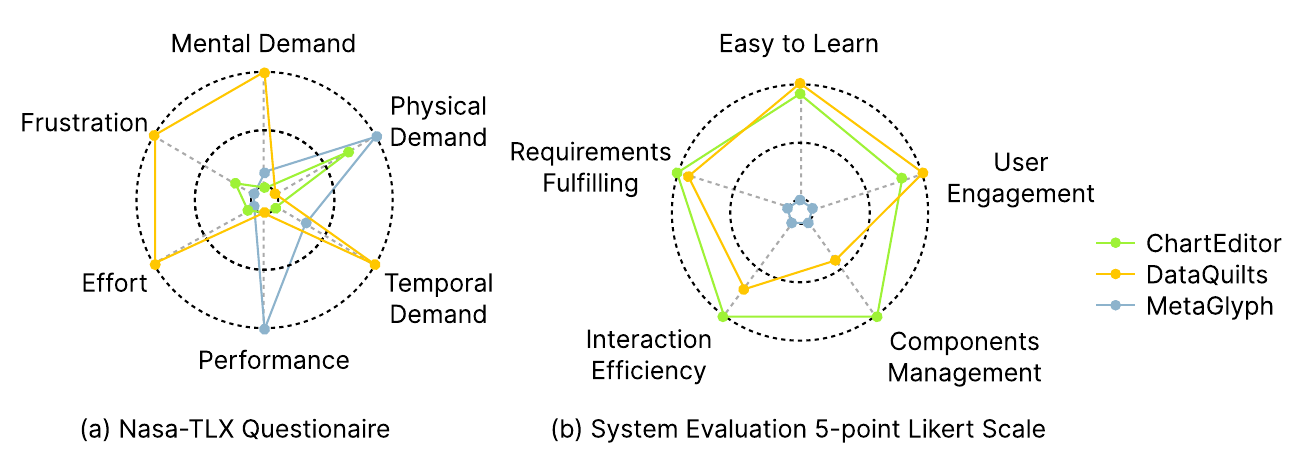}
    \caption{Radar chart presenting the comparative results of user study data for different chart authoring tools. \textbf{The data has been normalized, the purpose of which is to amplify the minor differences and make the contrast more obvious.} Please note that the Performance scale in panel (a) is inversely rated—a lower score indicates better performance. In other words, on this particular scale, the lower the score, the higher the perceived performance by the users, and vice versa.}
    \label{fig:radar}
\end{figure}

\stitle{NASA—TLX Questionnaire Results}. \cref{fig:radar}(a) illustrates the workload of various chart authoring tools across different aspects. The radar chart highlights that our method, \sys, performs exceptionally well, showing the lowest workload in both mental and temporal demand. It also demonstrates relatively low levels of effort and frustration. Interestingly, our method shows the highest physical demand among the three tools, while Metaglyph wins the lowest physical demand, which, however, consistently performs lower across most other indices compared to \sys.

% The overload evaluation of \sys revealed generally positive feedback across all categories, with mean scores ranging from 3.78 to 4.28. The highest-rated three categories were the error correction and undo functions (M = 4.28, SD = 0.67), Chart Tree functions (M = 4.22, SD = 0.65), and Interactive Design of the System Efficiency (M = 4.22, SD = 0.65). In contrast, the system's layout was rated slightly lower (M = 3.78, SD = 1.00), with higher variability, suggesting mixed opinions on its intuitiveness. An ANOVA test (F = 0.758, p = 0.640) showed no significant differences between categories, indicating that participants rated all aspects of the system similarly high. Pairwise t-tests further supported this finding, as no category pair demonstrated statistically significant differences (p > 0.05). 

\stitle{System Evaluation with 5-point Likert Scale}.
The systems are evaluated using a 5-point Likert scale, focusing on several key aspects derived from user feedback, as follows.

\begin{itemize}
\item \textit{Ease of Learning}: whether users appreciate its straightforward navigation and accessible functions;
\item \textit{User Engagement}: whether users enjoy its exploration; 
\item \textit{Components Management}: whether the CRUD operation is clear and effective;
\item \textit{Interaction Efficiency}: whether the system excels in efficiency and intuitiveness and reduces user effort.
\item \textit{Requirements Fulfillment}: whether its capability meets user requirements, earning top scores in this category.
\end{itemize}

As shown in \cref{fig:radar}(b), the results provide quite positive insight into the strengths and areas for improvement in our system.
Overall, the response to the scale was very positive. The system particularly excelled in interaction efficiency, components management, and fulfilling requirements
While ratings for ease of learning and user engagement were slightly lower, they were still favorable, indicating a balanced and user-friendly system.

\subsubsection{Qualitative Results}

This section provides a thematic analysis of the \sys based on audio scripts collected through a diverse source of think-aloud and semi-structured interviews.
% \sys is assessed across various aspects of user experience, with a focus on understanding its strengths and identifying potential areas for improvement. 
% This evaluation draws on insights regarding how the system supports user needs, facilitates creative tasks, and balances between automation and user control.
% The thematic analysis method~\cite{braun2006using} is employed to analyze the data collected from the semi-structured interviews and observation.
% The data was analyzed using open coding in a deductive, bottom-up approach by two researchers. The authors chose thematic analysis as it is well-suited for identifying recurring themes in participants' behaviors and viewpoints, making it an ideal method for examining user challenges with \sys. The interviews were independently coded by the two researchers using Freeform. Following this, they collaborated to discuss and compare their annotations, reaching a consensus to merge their codes into a unified set.

\stitle{Flexibility}. The system’s balance between simplicity and customization was well-received. Participants appreciated the ability to customize elements efficiently without being overwhelmed by excessive options. S6 commented, ``It feels just right to use — not too few options to be limiting, but not so many that it becomes overwhelming.'' S10 agreed, ``The way you adjust the charts is really straightforward. You don't have to go through a ton of steps to make changes, it's just right—nothing too complicated.'' Meanwhile, participants suggested that more advanced options could enhance the system’s flexibility. S17 grumbled, “I attempted to align several intricate visual elements, and the system struggled to provide precise feedback."

\stitle{AI-Driven Assistance}. % The AI-powered features were commended for their ability to streamline repetitive tasks and enhance productivity.
Participants appreciated the system’s capacity to handle vague inputs and generate diverse design output with AI tools. S7 mentioned that ``I wasn’t sure exactly what kind of style I wanted at first, but the AI made some good suggestions.'' However, they also pointed out 
the AI’s outputs
% occasionally lacked the refinement necessary for professional standards, 
often requiring manual adjustments. 
S16 suggested, ``You can consider adding a feature that allows me to upload custom elements or icons. This way, if the AI-generated result isn't accurate, I can step in and adjust it myself, rather than relying completely on the AI.''

\stitle{Quality of Generated Chart}. % The \sys 's templates and visual design tools were valued for their ability to simplify the creation of visually appealing designs. 
S3 praised pre-defined design patterns of the \sys, ``I’m really impressed with how good the generated charts look -- it’s like the tool does all the hard work, and I just get to tweak the final touches. I can’t wait to show off my charts!''
In addition to the ease of use, several participants mentioned that they enjoyed the process and found it fun to use the tool.
S6 said, ``It’s actually fun to create charts with this tool! It doesn’t feel like work, and I enjoy seeing the final result come together so quickly.''

\stitle{Ease-of-use}. Participants found \sys interface intuitive and easy to use; S2 appreciated that ``The icons were easy to understand, and I could figure out most of the features without much guidance.'' These elements streamlined navigation and enhanced the design process. S8 said, ``One-click auto-generation is so cool! As someone who is a bit lazy with design, it is perfect for me - it saves a lot of time.''
% \textbf{Workflow and User Experience:} 
% The workflow was described as intuitive and efficient, enabling participants to complete standard tasks with ease. The balance between automation and manual control was particularly appreciated, as it allowed users to quickly achieve their design goals while retaining the flexibility to make adjustments. Participants did, however, suggest that more granular control over certain aspects of the workflow could improve its effectiveness for complex projects.

\section{EXPERT FEEDBACK AND DISCUSSION}

In this section, we explore the development of chart authoring approaches and their future direction in the era of human-AI collaboration, guided by insights from a focus group of expert collaborators. During our user study, UX experts observed and recorded user interactions with the chart authoring tools. A subsequent 30-minute focus group captured feedback on user behaviors and experiences, highlighting the strengths of \sys while identifying limitations and opportunities for improvement.

\stitle{Flexible Hierarchy Design}.
The chart tree feature in \sys enables users to add, delete, view, and modify components at varying levels of granularity. Experts E1, E3, and E4 noted that traditional tools often rely on linear undo actions, leading to content loss when users attempt to edit specific components. The hierarchical approach, \ie \textit{chart tree}, in \sys addresses this by allowing precise adjustments without affecting other components, mitigating the unpredictable nature of AIGC. Inspired by layer-based systems in tools like Adobe Illustrator, this method enhances control and interaction efficiency, reducing the impact of AI's unpredictability on creative tasks~\cite{DBLP:journals/corr/abs-2411-01606}.
% Our design offers greater flexibility by incorporating a chart tree feature, allowing users to freely add, delete, view, and modify components at different granularity levels.
% This capability led to higher scores in interaction efficiency and component management during our evaluations. 
% Experts E1, E3, and E4 pointed out that in other chart authoring tools, when users are dissatisfied with previously generated components, they can only undo actions linearly back to the initial state.
% This limitation prevents them from individually editing the unsatisfactory components, often resulting in the loss of content that was already satisfactory.
% Due to AIGC's uncontrollable nature, once the original parts are undone, even using the same prompt rarely reproduces the same content.
% We believe that representing components hierarchically has a positive impact on chart authoring approaches.
% This method integrates layers similar to those in Adobe Illustrator and Photoshop.
% Since the unpredictability of AI-generated content is an urgent issue, our approach helps reduce its negative impact on tasks by providing users with more precise control over individual components.

\stitle{Intent Formalization}.
\sys bridges the gap between vague user intentions and actionable outcomes through its natural language prompting system, which formalizes user intent and reduces cognitive load~\cite{10720675,DBLP:conf/sigmod/Luo00CLQ21,DBLP:journals/tvcg/LuoTLTCQ22}. Experts E2 and E4 observed that users often paused or struggled with other tools due to unclear workflows, whereas \sys streamlined their thought processes. Expert E3 highlighted that users frequently switched to \sys to refine outputs before returning to their original tools, underscoring the value of intent formalization in enhancing workflow efficiency and achieving desired outcomes faster.
% Prompting helps formalize users' intent, which enhances the accuracy and effectiveness of automatic pictorial chart generation systems. It bridges the gap between vague or high-level user intentions and specific, actionable modifications.
% Experts E2 and E4 observed that users often pause or become lost in thought when working with other tools and sometimes forget to think aloud.
% This suggests that idea prompting helps reduce cognitive load by minimizing the mental effort needed to achieve goals, ultimately enhancing workflow efficiency. 
% Expert E3 noticed that some users, when using a certain tool, always switch to \sys for authoring and then return to the original tool.
% This behavior suggests that the lack of intent formalization in other tools is a key limitation.
% Users prefer tools like \sys that can translate their ideas into executable outputs, enabling them to complete complex creative tasks more efficiently and achieve their desired results faster.

\stitle{Pictorial Charts Generation Based on Basic Charts}. 
\sys focuses on transforming basic charts into pictorial versions by replacing standard elements with contextually relevant visuals, preserving data integrity while enhancing appeal. An alternative approach, \ie generating pictorial charts directly from raw tabular data, could simplify the process for users lacking basic charts. However, this method introduces challenges such as inconsistent visualizations due to the absence of structural constraints. Future work will explore this approach, aiming to develop models capable of interpreting data context and structure to produce clear and intuitive pictorial charts.

% This method focuses on transforming basic charts into pictorial versions by replacing standard elements like bars or lines with contextually relevant objects. This preserves data integrity while enhancing visual appeal. 
% An alternative approach is {\em generating pictorial charts directly from raw tabular data}, bypassing the need for a basic chart. This method simplifies the process for users without pre-existing basic charts and offers more creative flexibility, allowing the system to generate a chart from scratch.
% However, without the structural constraints of a basic chart, it risks producing inconsistent or unclear visualizations, which can make understanding these pictorial charts more difficult.
% Future work will explore direct pictorial chart generation from raw data, focusing on developing models capable of interpreting data structure and context to create accurate and intuitive pictorial charts.

\stitle{Extending the Functionalities of ChartEditor}.
Enhancing \sys with advanced customization options could expand its utility. Features like user-defined icon libraries, more detailed styling options (e.g., borders, gradients, animations), and context-based design suggestions would cater to diverse user needs. Intelligent recommendations for icons, colors, or layouts could assist non-designers in maintaining both accuracy and aesthetic quality, making \sys more versatile for various use cases.
% One area for extension involves advanced customization options. While \sys provides a streamlined process for transforming basic charts, users would benefit from the ability to personalize their charts further. For example, integrating user-defined icon libraries would allow users to upload their own icon sets, enabling them to create charts that align with specific branding or thematic requirements.
% In addition, providing more control over the chart components enables users to customize the charts and match personal preferences.
% While \sys allows for basic customization, future versions could include features for advanced styling akin to those found in tools like Microsoft PowerPoint. This could include options for adding borders, shadows, and gradients or allowing for more intricate animations and transitions between chart elements.
% Moreover, context-based design suggestions could significantly enhance the user experience. Specifically, \sys could offer intelligent suggestions for visual elements, such as recommending appropriate icons, colors, or chart layouts, by analyzing the data and the type of chart being created. Such a feature would also cater to non-designers by offering guidance that ensures both data accuracy and aesthetic quality.

\stitle{Supporting More Chart Types}.
Currently, \sys supports basic chart types like bar, line, and pie charts. Expanding its repertoire to include more complex types, such as heatmaps, scatterplots, and treemaps, would broaden its applicability across diverse industries and tasks, enabling richer data visualization capabilities~\cite{DBLP:conf/emnlp/WuYSW0L24}.
% Currently, \sys supports a limited range of basic chart types, including bar, line, and pie charts. While these types are widely used and effective for common data visualization tasks, expanding \sys to accommodate more complex chart types such as heatmaps, scatterplots, and treemaps would significantly enhance its applicability across a broader range of use cases and industries.

\section{Conclusion}
In this paper, we presented \sys, a human-AI paired tool that allows users to create pictorial charts from basic charts through natural language interaction.
To enable users with precise control over chart elements, we introduced the chart tree, a hierarchical structure that organizes chart components for efficient modification and editing.
To facilitate the decomposition of chart images and integrate them into the chart tree, we curated a large-scale dataset, ChartSS, and fine-tuned a chart segmentation model specifically for this task. 
Finally, we conducted user studies that demonstrated \sys's usability and effectiveness on the task of pictorial chart generation through a combination of AI-driven automation and user-guided refinement.

\bibliographystyle{IEEEtran}
\bibliography{reference}

\begin{IEEEbiography}[{\includegraphics[width=1in,height=1.25in,clip,keepaspectratio]{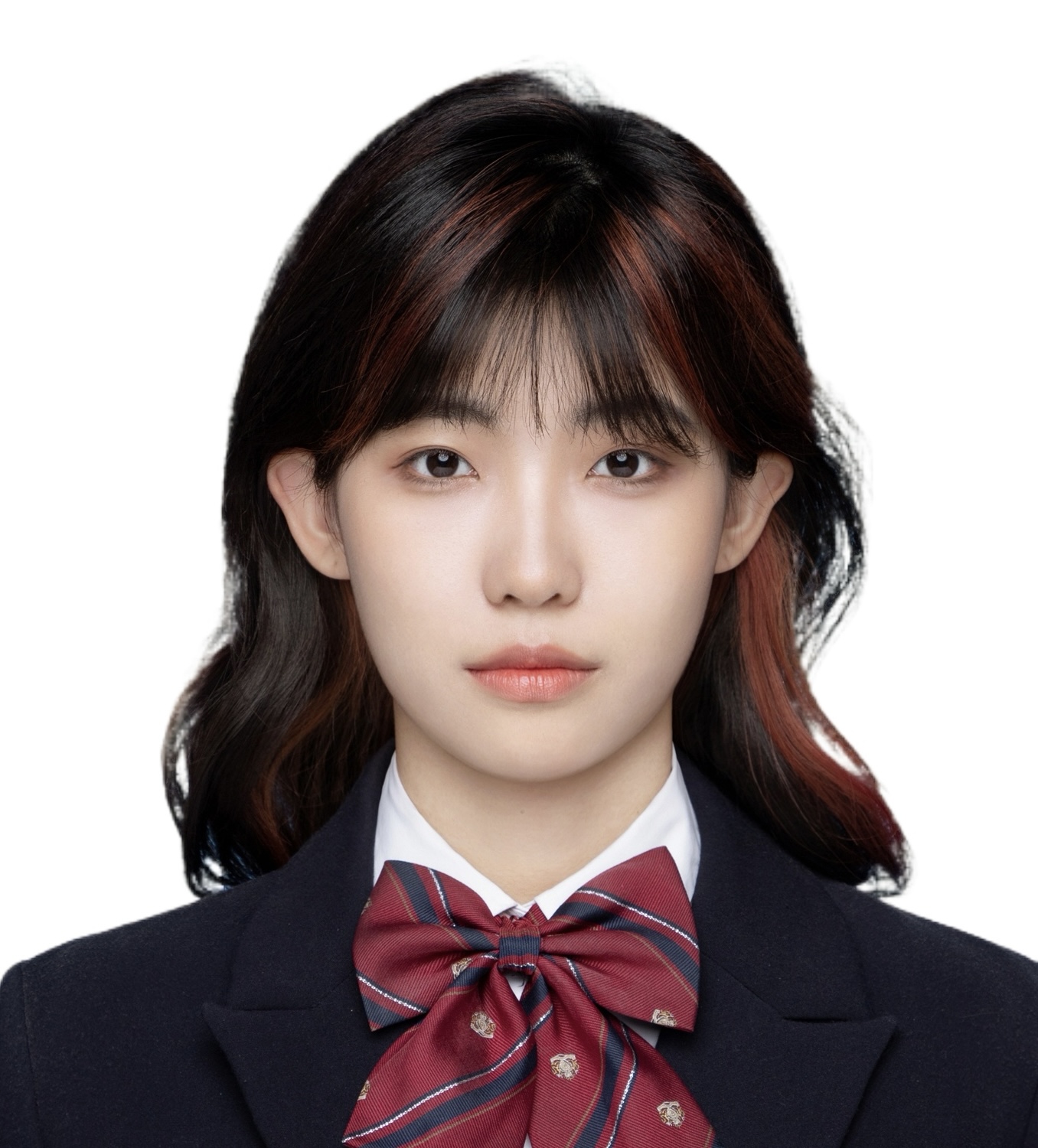}}]{Siyu YAN}
 is an MPhil student at the Info Hub, Hong Kong University of Science and Technology (Guangzhou). She received her bachelor's degree from Nankai University. Her current research focuses on AIGC and AI-driven Data Analytics.
\end{IEEEbiography}

\begin{IEEEbiography}[{\includegraphics[width=1in,height=1.25in,clip,keepaspectratio]{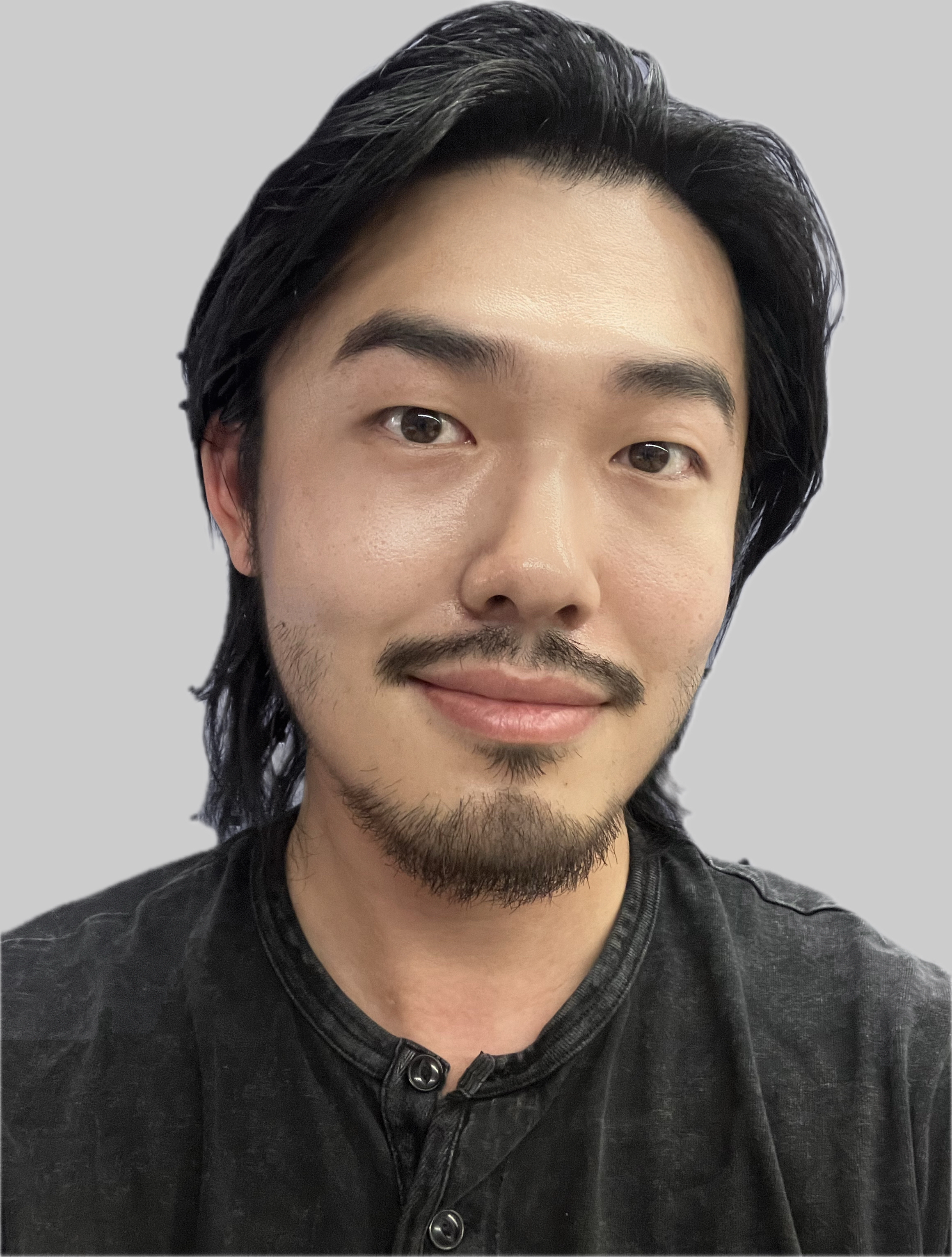}}]{Tiancheng LIU}
 is a Ph.D. candidate at the Info Hub, Hong Kong University of Science and Technology (Guangzhou). His current research focuses on cultural heritage, computing aesthetics, and their application in multimodal large language models. He also has experience in blockchain technology, cloud networking, and IoT engineering, as well as intelligent traffic systems management and data-centric AI.
\end{IEEEbiography}

\begin{IEEEbiography}
[{\includegraphics[width=1in,height=1.25in,clip,keepaspectratio]{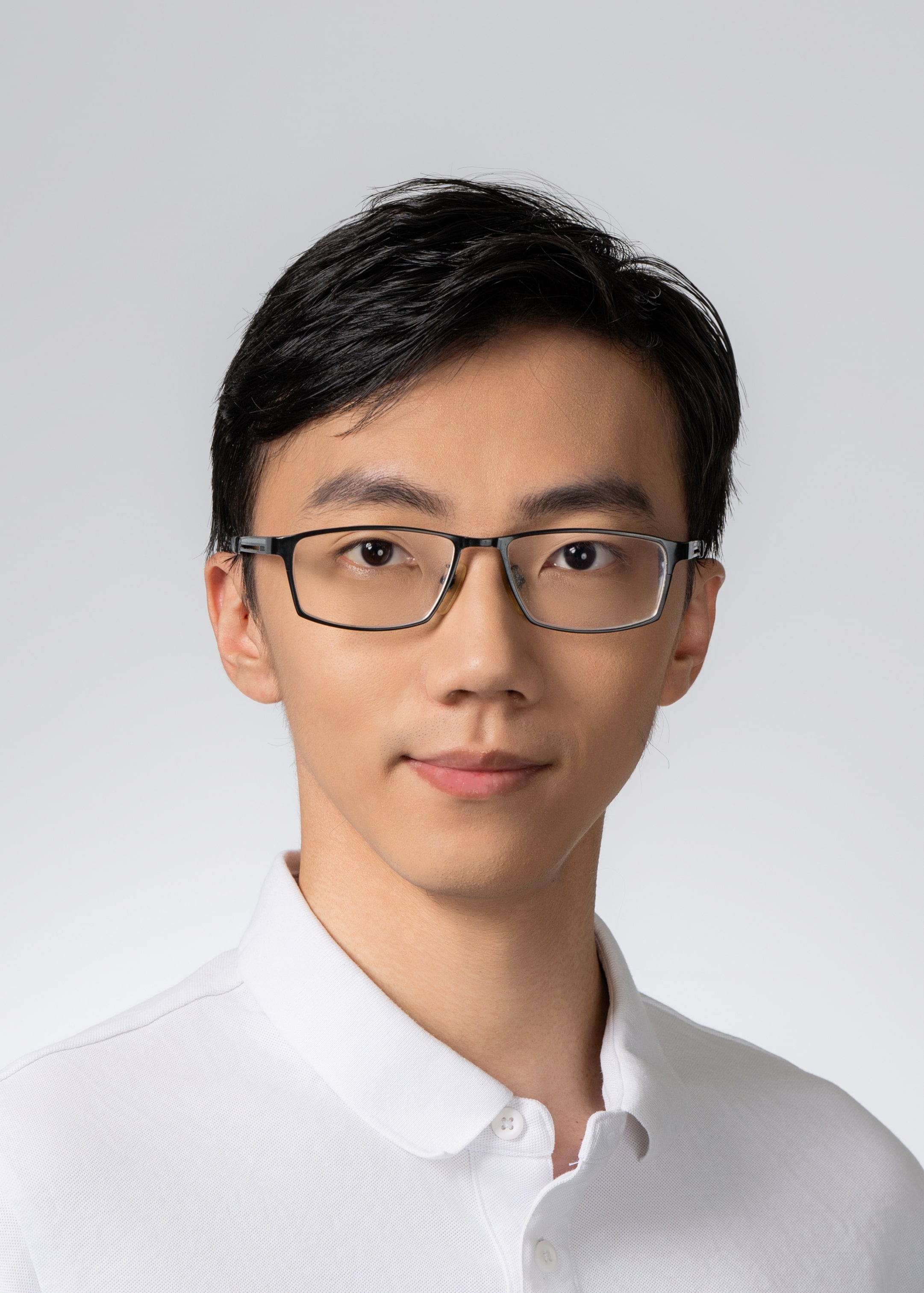}}]{{Weikai Yang}} is an assistant professor in Hong Kong University of Science and Technology (Guangzhou). His research interests lie in visual analytics, machine learning, and data quality improvement. He received a B.S. and a Ph.D from Tsinghua University.\looseness=-1
\end{IEEEbiography}

\begin{IEEEbiography}[{\includegraphics[width=1in,height=1.25in,clip,keepaspectratio]{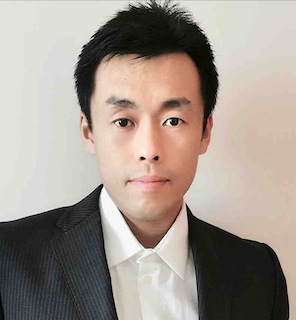}}]{Nan Tang}
 is an associate professor at the Hong Kong University of Science and Technology (Guangzhou), and is affiliated with the HKUST. He has received the VLDB 2010 Best Paper Award, the 2023 SIGMOD Research Highlight Award, and the SIGMOD 2023 Best Papers. Dr Nan’s main research interests are data management, visual analytics, and data-centric AI.
\end{IEEEbiography}

\begin{IEEEbiography}[{\includegraphics[width=1in,height=1.25in,clip,keepaspectratio]{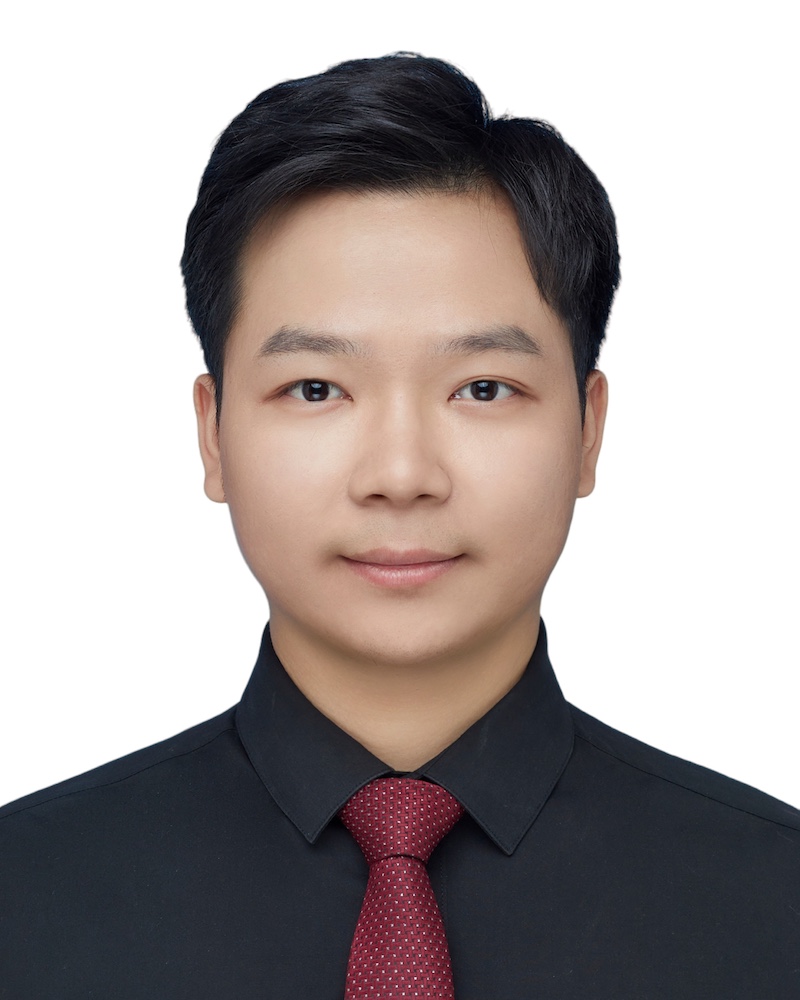}}]{Yuyu Luo} is an assistant professor at The Hong Kong University of Science and Technology (Guangzhou), with an affiliated position at the HKUST. He received his PhD from Tsinghua University in 2023. His research interests include AI-driven data analytics, visual analytics, and data-centric AI. He has received the SIGMOD 2023 Best Papers.
\end{IEEEbiography}

\vfill

\end{document}